\documentclass[12pt,preprint,times]{aastex}

\slugcomment{to appear in AJ}

\shorttitle{The Stellar Populations of NGC 185 and NGC 205}
\shortauthors{Butler \& Mart\'{\i}nez-Delgado}

\begin{document}
\title{On the stellar
   populations in  NGC 185 and NGC 205, 
 and the nuclear star cluster in NGC 205
    from {\it Hubble Space Telescope}\thanks{Based on
  observations made with the 
  NASA/ESA Hubble Space Telescope, obtained from the data archive at the Space
  Telescope Science Institute. STScI is operated by the Association of
  Universities for Research in Astronomy, Inc. under NASA contract NAS
  5-26555.} observations} 
\author{D. J. Butler \& D. Mart\'{\i}nez-Delgado}
\affil{Max-Planck-Institut f\"ur Astronomie, K\"onigstuhl 17, D-69117
  Heidelberg, Germany}
\email{butler@mpia.de, ddelgado@mpia.de}

\begin{abstract}
   We present the first detailed analysis of resolved stellar
 populations in the dwarf galaxies  NGC 185 and NGC 205 
  based on  archival V- and I-band WFPC2 pointings.

 For NGC 185 we  deduce from the brightest main sequence and 
 blue loop stars that 
   star formation  was probably still
 active about  4$\times$10$^8$\,yr ago; and  
  have three key abundance-related results: (1) We  identify
 ancient  stars with  [Fe/H] $\la$ -1.5\,dex by a well-defined
 horizontal branch (HB). 
 (2) We find a prominent clump/bump-like feature along the red giant 
branch/faint asymptotic giant branch (RGB/faint AGB) with the same mean 
V-band magnitude as in the HB, within uncertainties (i.e.,
 $\Delta$V(Bump - HB) = 0); from 
 a comparison with theory,  the implication is that
 ancient stars have [Fe/H] $\sim$ -1.5\,dex,  
 with a  higher abundance level for intermediate-age stars.
  (3) From colour information we infer that median [Fe/H] $>$
 -1.11$\pm$0.08\,dex for ancient stars  (assuming E(B-V) = 0.18\,mag). 
   
 For NGC 205, we record a new distance modulus, (m-M)$_{\rm 0}$ =
 24.76$\pm$0.1\,mag,  taking E(B-V) = 0.11\,mag,
 based on the red giant branch (RGB) tip magnitude method in I-band.
 We find that stars were probably still forming  less than  
 3$\times$10$^8$\,yr ago in NGC 205, which  is  compatible with 
 star formation triggered by an interaction with M31.
 Three key abundance-related results for NGC 205 
 are: 
 (1)  The red giant/faint-asymptotic giant branch is significantly skewed  to 
  redder values than that of a  
 control field in the outskirts of M31;
  it probably results from 
 a relatively narrow metallicity and or age range for a significant 
 fraction of the dwarf's stars.
 (2) From a comparison with models, the most metal-rich RGB
   stars  reach 
  to  [Fe/H] $\ga$ -0.7\,dex ($\ga$0.2\,Z$_\odot$). 
    (3) For  ancient stars  we infer from
 colour information that median [Fe/H] $>$
  -1.06$\pm$0.04\,dex (for E(B-V) = 0.11\,mag). 
    We briefly compare the stellar populations of  NGC 205,
  NGC 185 and NGC 147.  
 
 Finally, we  study  several  V- and R-band 
 structural properties of the nuclear star 
 cluster in  NGC 205 for the first time: The  apparent  V-/R-band 
 effective radii  indicate 
   a blue excess in  the  cluster's
  outer region. In terms of size, the cluster is 
 like a typical galactic globular cluster or  a nuclear cluster 
 in a nearby late-type spiral galaxy; but it is quite bright
 (10$^6$\,L$_{\rm \odot,R}$), unlike an ancient globular cluster,
  and by matching with models, the blue
  colour  hints that its stellar population is young, up to a few
 times 10$^8$\,yr old. 
\end{abstract}

\section{Introduction}\label{intro}
 Successive merging and 
 accretion of many comparatively small stellar systems  
 like dwarf galaxies, as well as intergalactic gas, 
 is an important part of the theoretical framework in which 
 large (disk) galaxies are predicted to have formed 
 (e.g., White \& Rees 1978; Steinmetz \& Muller 1995). 
 Significantly, studies of 
  resolved stars in dwarf companions ultimately allow one to explore two paired
 issues, namely giant galaxy assembly as well as 
 the star formation history of dwarfs,
  which describes the gradual conversion 
 of gas into stars, followed by
  the expulsion of remaining gas by stellar winds 
 and the resultant heating and
  cooling of the interstellar medium.
 
 Since Baade (1944a, 1944b) first succeeded  in resolving the
 brightest stars of  several dwarf elliptical (dE) satellites of the giant spiral galaxy Andromeda, 
 those galaxies have been regarded as simple, old stellar systems,
 with a stellar content  resembling that
 of Galactic globular clusters (Population II). From the mid-1990s onward
  however
  the   impressive improvement in the quality of 
  colour-magnitude diagrams (CMDs) for nearby
 dE galaxies provided by the HST has revealed that those ``simple"
 galaxies often display varied, and in some cases, complex star formation
 histories (SFHs)  (Da Costa 1998; Mateo 1998).  However, it remains unclear
 whether the  Andromeda dE companions
 (NGC 205, NGC 147 and NGC 185) have had significant  star formation   
 episodes since the  primeval event
 (see Da Costa 1998 for a recent review). 

   The existence of a dozen luminous blue stars (Baade 1944a) and other   
 Population~I features (dust clouds, HI gas, SN remnant) at the centre
 of NGC 185 has been an intriguing feature of this galaxy for several
 decades because such galaxy building blocks had been alledged to have
 been depleted at earlier formation epochs in elliptical  galaxies.
   This  recent star formation epoch is confined to its central
 150 $\times$ 90\,pc, where the youngest, 100\,Myr old, population is
 found (Hodge 1963; Lee, Freedman \& Madore 1993; Martinez-Delgado,  
 Aparicio \& Gallart 1999, hereafter MD99). 
 In NGC 147 however, there are no signs of recent star formation ( in
 the past 1\,Gyr) or  gas at its centre   (Han  et al. 1997;
 Young \& Lo 1997; Sage, Welch \& Mitchell 1998).
 Thus, this difference 
 between NGC 185 and NGC 147 is intriguing because otherwise they are 
 similar in many of their properties (e.g. type, mass,
 size).
 Similarly, NGC 205 has been regarded as an unusual dE   galaxy
 -- it has  gas  (Johnson \& Gottesman 1983; Young \& Lo 1997) and a 
 young central stellar population  
 (Baade 1944b; Lee 1996), and many of the brightest 
 objects have recently been resolved into groups of UV-bright 
 stars (Cappellari et al. 1999). 
 So, why do 
 the recent ($<$ 1\,Gyr) SFHs of Andromeda's dE companions differ ?

 A natural way to probe this issue is to 
  search for an intermediate-age (1-10 Gyr)
 population in these galaxies. The first evidence of such a population
 in these galaxies was the presence of stars above the tip of the red giant
  branch  (TRGB) in CMDs from ground-based observations
  of their inner regions (NGC 205: Lee 1996; NGC 185: Lee  et al. 1993,
 Martinez-Delgado \& Aparicio 1998, hereafter MDA98;  
 and NGC 147: Davidge 1994) and have been interpreted as intermediate-age
 AGB stars by several authors (Davidge 1994;
 Lee et al. 1993; Lee 1996).  However, ground-based observations 
 can lead to severe central crowding, making RGB stars appear brighter than
 they are,  mimicking an intermediate-age AGB population (Martinez-Delgado \&
 Aparicio  1997). 

 Two other paired issues are the origin and possible stripping
 of the gas from which  
 young stars are born in such galaxies. One could view the
  concentration of young stars
 lying in the central part  of NGC 185 and NGC 205 
  as a hint that new stars
 could originate from material ejected by dying stars
 and that this process would only be efficient enough at the center   
 of the galaxies.
 This would suggest an internal origin for the gas in these
 systems, as was argued by Faber \& Gallagher (1976). However,
 Young \& Lo (1997) find different scenarios for the three  Andromeda
 dE satellites: no gas in NGC 147 
 (M$_{\rm HI}$/L$_{\rm V} \la 4 \times
 10^{-5}$M$_\odot$/L$_\odot$)\footnote{M$_{\rm HI}$/L$_{\rm V}$ data is taken
 from from Grebel,
 Gallagher \& Harbeck (2003; their Table 1)};
 gas with a  different angular
 momentum per unit mass  to the stars in NGC 205
 (M$_{\rm HI}$/L$_{\rm V} \sim 1 \times 10^{-3}$M$_\odot$/L$_\odot$); 
 and gas kinematically compatible with
 that of the stellar component in NGC 185 
 (M$_{\rm HI}$/L$_{\rm V} \sim  1 \times 10^{-3}$M$_\odot$/L$_\odot$).
  Although several theories (e.g. external origin: Knapp  et al. 1985;
 ram-pressure stripping: Lin \& Faber 1983; or  
 different initial conditions
 -- (e.g., gas number density and gas mass): Hensler \& Burkert 1989)
  have been proposed to explain
 this puzzling situation in dwarf galaxies, 
 the exact origin of the current inner region differences 
  remains a mystery.

 In the present paper, we draw on   
 a new colour magnitude diagram--based analysis of 
 resolved stellar populations in NGC 205 and NGC 185 
 to take a new look at their  ages and 
  abundances.
  We re-analyze  NGC 147 data for comparative purposes.
 
 The layout of the present paper is as follows.
 The data and data reduction are described in
  Sec.~\ref{obs_datared}. Extinction is assessed briefly in
 Sec.~\ref{Sec_extinction} and an overview of the stellar content
 based on an inspection of our 
 colour magnitude diagrams is given in Sec.~\ref{CMD_description}.
  We infer the distance modulus and [Fe/H]
   in  Sec.~\ref{Sec_dist_metal}.  
 Abundance distributions are probed in Sec.~\ref{Secabundance_distrib},
 and we make a comparison with previous studies and a control field  in 
 Sec.~\ref{tails}. In Sec.~\ref{redclump}, 
 we briefly discuss the possible nature of 
 the red clump/bump feature in NGC 185.
 Results concerning  young-to-intermediate--age stars 
 are reported in Sec.~\ref{youngstars}.
  In addition, as NGC 205 offers an opportunity to 
 study a bona fide nucleated  dwarf satellite in our neighbourhood 
 we  report several newly studied 
 properties of its central compact star cluster 
 and  comment briefly on NGC 205 
 in the context of tidal stripping (Sec.~\ref{NC_SBfitting}).  
 Finally, we 
 discuss briefly the nature of NGC 205
 in Sec.~\ref{stripping} and 
 summarize our main conclusions  in  Sec.~\ref{Conclusions}.

\section{Data and data reduction}\label{obs_datared}

\begin{figure*}\label{chart_ngc185}
\includegraphics[width=150mm,height=65mm,angle=0]{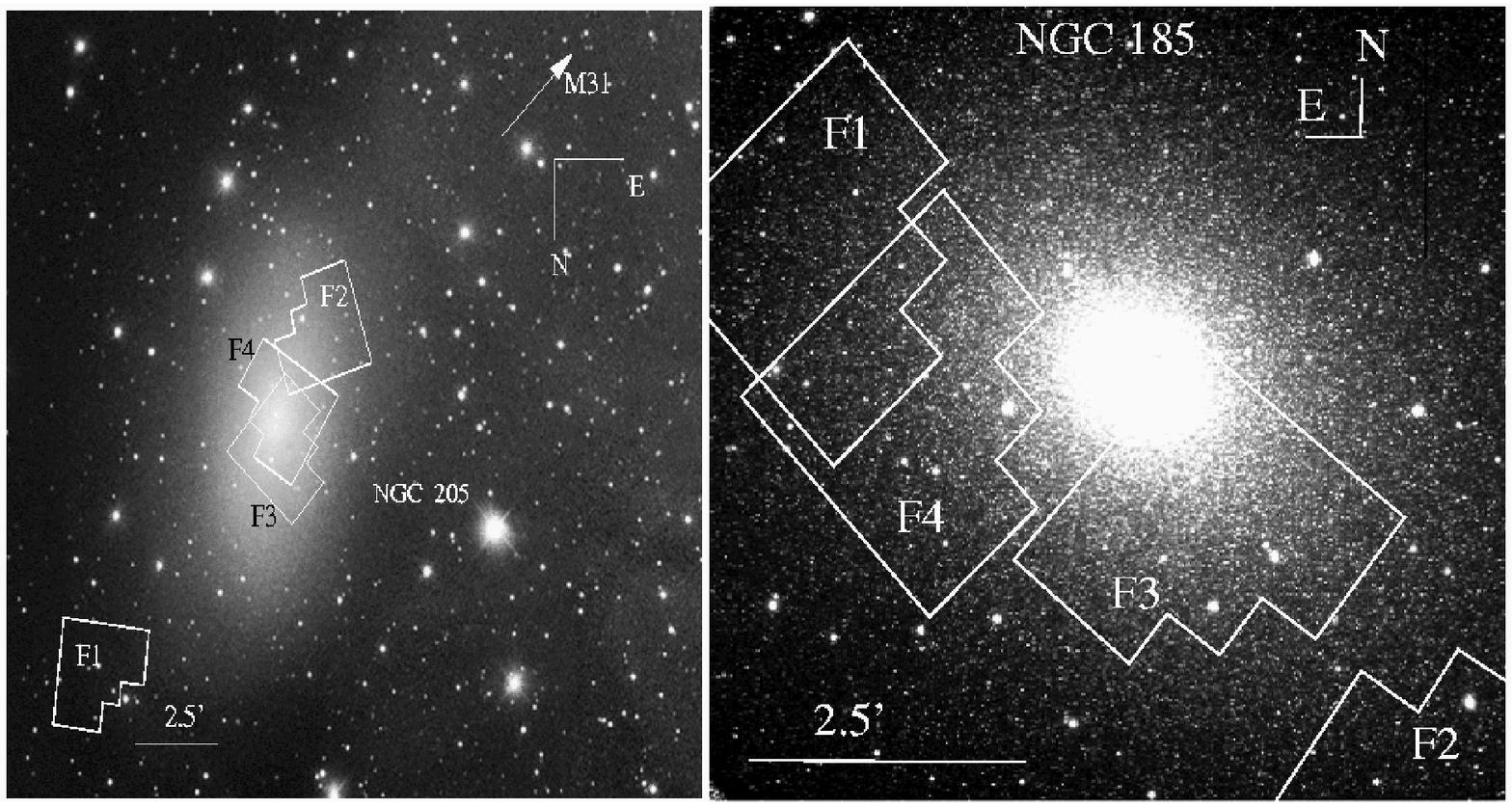}
\caption{False optical colour mosaic 
 image of NGC 205  (left; Credit: R. Gendler (2002), 0.32\,m telescope, F/6).
  I-band image of NGC 185 (right; Nowothny et al. 2003).  Locations of the  archival pointings
 are indicated with full or partial WFPC2 footprints. }
\end{figure*}

 In the present study of NGC 185 and NGC 205 (and NGC 147),
 calibrated science- and data-quality images have been retrieved from the Space
 Telescope  Science Institute HST data archive. I (F814W)- and V (F555W)-band frames from  selected HST/WFPC2 
 pointings has been examined.  
 For our study of the nuclear star  cluster
 in NGC 205 we  retrieved V(F555W)- and R (F675W)-band frames.
 Source information is given  in Table~\ref{obs_log}; and relative positions
 of   WFPC2 
 footprints can be seen in Fig.~\ref{chart_ngc185}.

 For the data reduction, the following
 strategy has been applied. 
 Photometry of the stars  was determined using the
 HSTphot (Dolphin 2000a)  point spread function (PSF)-fitting 
 photometry package which  is  designed for optimal reduction
 and analysis of WFPC2 data. 
 Stellar magnitudes are calibrated by HSTphot using the charge 
transfer efficiency and zero
 point magnitude corrections derived by Dolphin (2000b) and 
 are transformed into the  V- and I-band passes.
 Lastly, artificial star tests 
 identical to those described in Butler, 
 Martinez-Delgado \& Brandner (2004) have been performed using HSTphot.

\begin{figure*}
\includegraphics[width=17.8cm,height=8.5cm,angle=0]{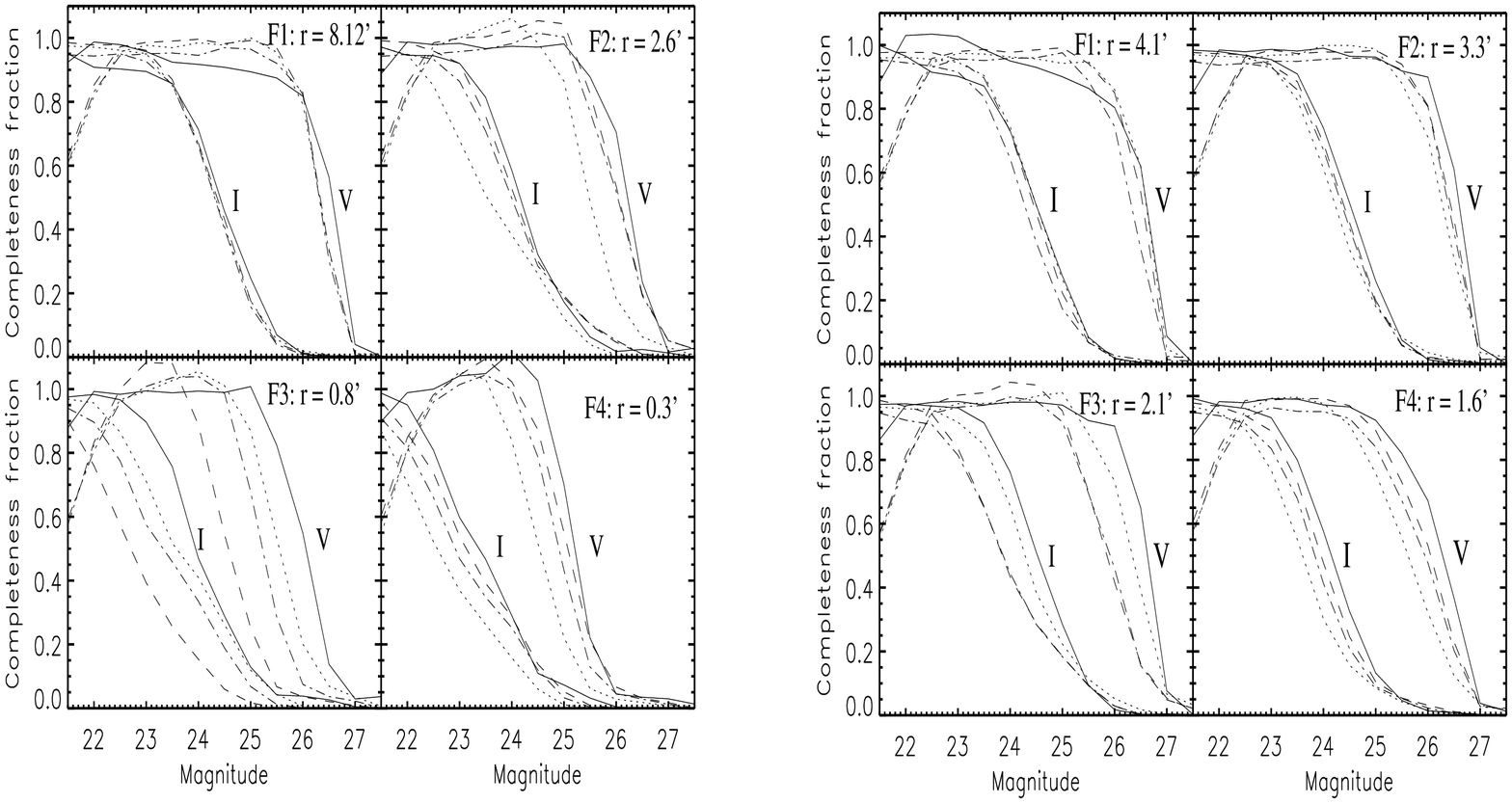} 
 \caption{V-, and I-band fractional completeness 
 as a function of magnitude 
 for WFPC-2   V-,  and I-band   PC1 ({\it solid}), WF2 ({\it dotted}), WF3
 ({\it dashed}) and WF4 ({\it dot-dashed})
 frames  at each pointing in NGC 205 (left) and NGC 185 (right)
 as determined  by artificial star tests. 
 See the text in Sec.~\ref{obs_datared} for further explanations.} \label{FigCompltnss1}
    \end{figure*}

 Fig.~\ref{FigCompltnss1}  plots the completeness
 function  against magnitude  for each pointing: The 
 completeness functions are the number of artificial stars recovered
  divided by the number of input stars.
 A completeness plateau generally 
 occurs for the brightest stars because their
  signal-to-noise 
  is sufficiently high and crowding affects these stars equally. 
 A completeness of  100\% is typically 
 not reached because of bad pixels\footnote {The
 number 
 of pixels deemed to be bad 
  in the unvignetted portion  of each chip/frame is of the order of 1\% 
     (includes cosmic ray events;  also includes columns and pixels flagged as 
 bad in the STScI data quality file).}. 
 In those frames
 containing one or more prominent star clusters, 
 an over-detection of bright stars
 at  bright magnitudes occurs due to the so-called
 bin-shifting effect -- 
 some faint artificial  stars have blended with 
 stars from the star cluster(s), and have been detected as  brighter stars.

For the final photometry list used in this study, objects have been selected
 if flagged as valid stars for both band-passes and have  S/N $>$ 5;
$\chi^2$ $<$ 7; -0.5 $<$ sharpness $<$ 0.5; and $\sigma_{\rm V, I}$ $<$
0.25\,mag.  
 The number of stars in each
 field after selection is given in Table~\ref{obs_log}.

\section{Extinction}\label{Sec_extinction}
 Knowledge of the extinction towards stars is a basic necessity
 for determining  their distances, masses,  chemical compositions and ages.
 For an upper limit on the extinction 
  from  M31's dust clouds in front of NGC 205, plus foreground extinction  
 due to the Galaxy,  we recall that extinction
  in several outer disk fields in M31 has been estimated to be
 of the order of E(B-V)= 0.13\,mag (e.g., Hodge \& Lee 1988;  Bellazzini et al. 2003  / BCF03)
  with a dispersion in their values of about  25\%. 
  For an estimate of the foreground extinction alone, towards M31/NGC 205, 
 we adopt E(B-V) = 0.086\,mag from the Schlegel, Finkbeiner \& Davis (1998;
 SF98) dust maps.\footnote{
M31 has strong far-IR emission in its spiral arms, which could potentially 
 contaminate a dust map, but it was excised 
 before SFD98 generated their final dust map.} 
 As NGC 205 is behind M31 (e.g. Sec.~\ref{Sec_dist_results}) with 
  a projected location at the far outer edge of M31,
 one might presume  E(B-V) $<$  0.13\,mag, i.e., 
  smaller than values measured using
  outer disk stars of M31. As a reasonable compromise, 
 we adopt the average of extinction data for the M31 outer disk 
 and the  SFD98 value, namely E(B-V) = 0.11\,mag, 
 and assume the Galactic reddening law with A$_{\rm V}$ =
 3.1$\times$E(B-V)  and A$_{\rm I}$/A$_{\rm V}$ = 0.48 
(Cardelli, Clayton \& Mathias 1989).

 A further issue is the  internal extinction in NGC 205.
 About a dozen dust patches  in the central 1-2$^{\prime}$ have been
  catalogued,  
 the largest of which points toward differential 
 extinction values of up to about 0.4\,mag  in V-band (Hodge 1973). Those
 patches occupy about 5\% of the F4 field,
  and their effect on results in the present paper is minor as determined
 by repeated tests in which photometry from
 5\% of each field was dimmed, conservatively, 
 by 0.3\,mag -- the resulting systematic
 change in RGB/faint-AGB colours is up to 0.02\,mag or 0.01\,mag in distance
 modulus with a negligible change in [Fe/H].
 However,  there remains the possibility of systematic errors with 
 a similar order of magnitude due to a larger area 
 being affected by a smaller level of extinction (e.g. 0.05\,mag).

  For NGC 185 and NGC 147,  foreground extinction estimates 
   from  SFD98 are  E(B-V) =  0.184\,mag  and 0.173\,mag
  respectively.  
 However,  NGC 185 and NGC 147 are located closer to the galactic plane 
 (b$\sim$-14.5$^\circ$) than NGC 205 (b$\sim$-21$^\circ$), 
  and so, it is important to be aware of the possibility of significant
 inaccuracy in SFD98 extinction estimates towards low
 galactic latitudes that has yet to be  established (for $|$b$|$
   $<$ 10$^\circ$ at least; see SFD98). 
 It is however rather encouraging that the above-mentioned extinction 
 estimate from SFD98 for NGC 185 is in very good agreement with the result of 
 a direct measurement of the stellar colour excess, i.e.  E(B-V) =
 0.19$\pm$0.03\,mag (Lee et al. 1993). 
  Extinction data from SFD98 
 is given for each galaxy in Table~\ref{Extinction1}.

\section{Stellar content from colour-magnitude diagrams}\label{CMD_description}
\begin{figure*}
\includegraphics[width=185mm,height=120mm,angle=0]{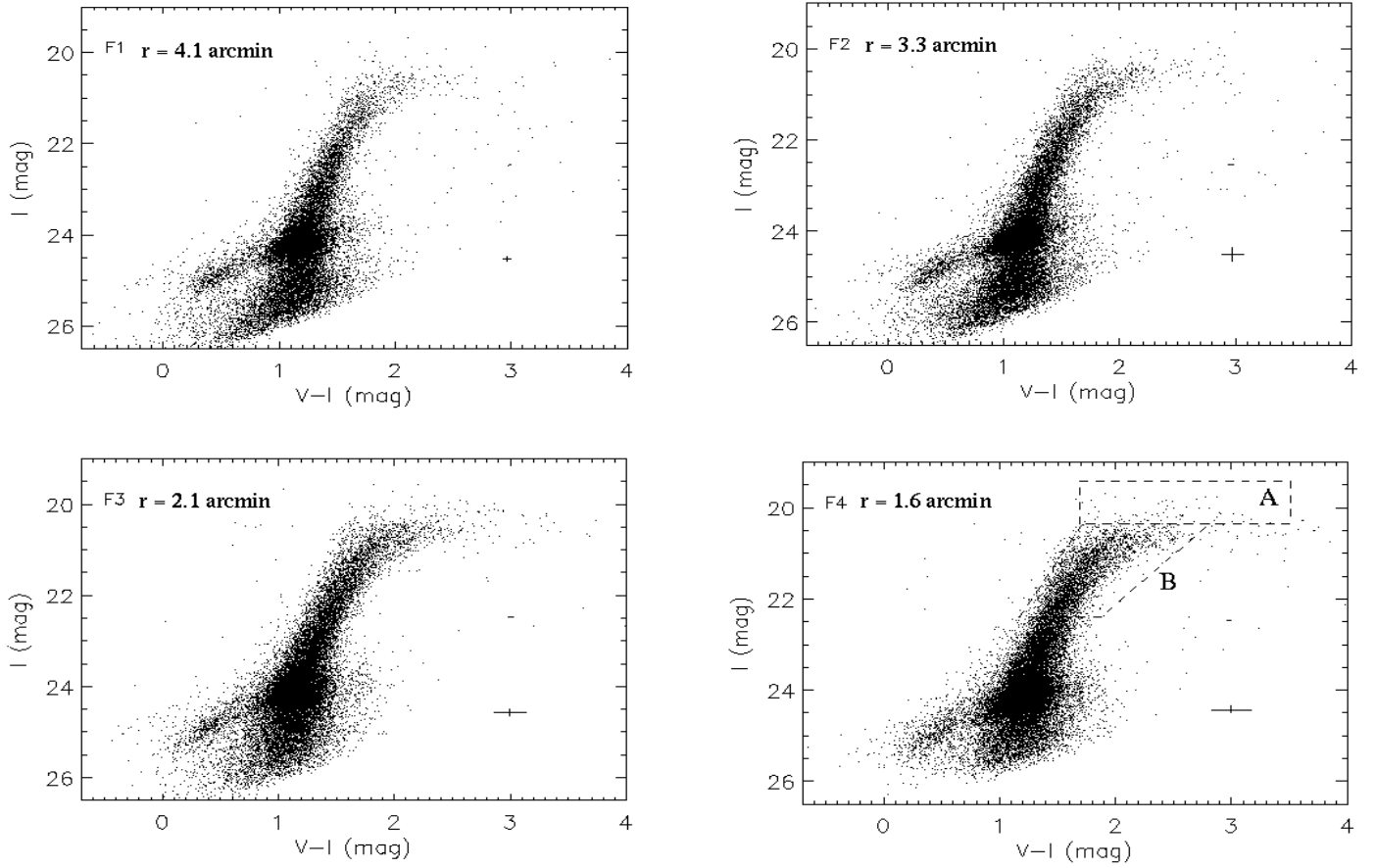}\\
 \caption{I, V-I CMDs for NGC 185.  (Median) error bars in magnitude and
  colour, taken at V - I = 1\,mag, are included.
  Box boundaries used for star counts given in Table~\ref{AGB_RGBratios}
  are shown for the field F4; I-band box boundaries are  M$_{\rm I}$ = -5.0,
  -4.06, and -2.0\,mag.  For further explanations see the text in   
 Sec.~\ref{obs_datared}. 
}\label{cmds1}
\end{figure*}

\begin{figure*}
  \includegraphics[width=185mm,height=120mm,angle=0]{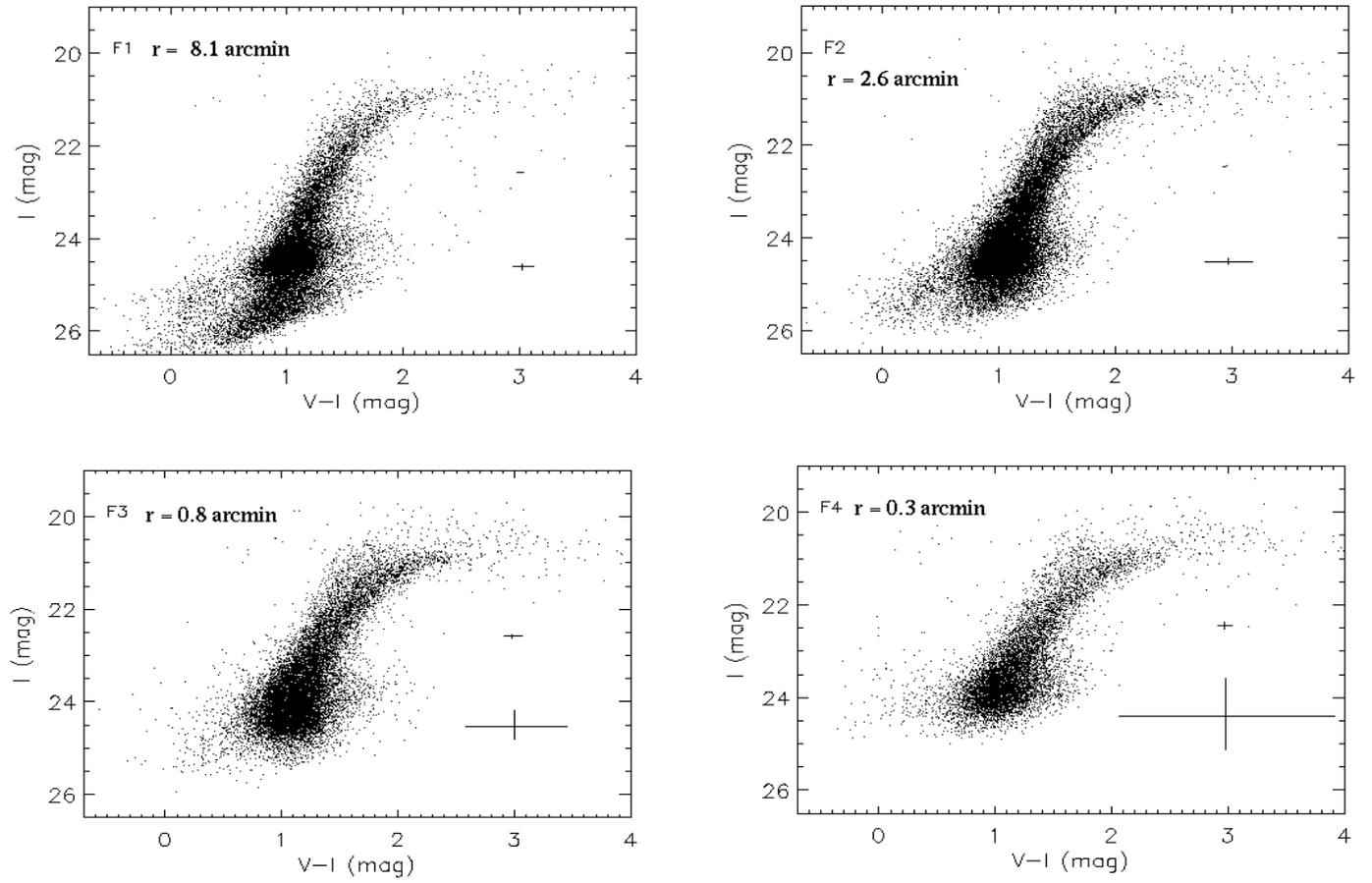}\\
\caption{As Fig.~\ref{cmds1}, except for NGC 205. }\label{cmds2}
\end{figure*}

 Figs.~\ref{cmds1} and ~\ref{cmds2} show  the (I, V-I) CMD  
 at each pointing in  NGC 185 and  NGC 205 respectively.
  They  display the brightest
 5-6\,mag of the RGB/faint-AGB sequence
  and  give information on the stellar content therein.

 In NGC 185, the  fields in general   look very similar.
The  most prominent feature is the RGB-like structure,
  the upper two  magnitudes of which is very
similar to that found by MDA98 in a low spatial resolution study of NGC 185. 
This structure is typical of a galaxy that  clearly comprises a mixed
 stellar population: It is the locus of old
 and intermediate-age RGB stars,  as well as faint AGB
 stars. There is a prominent red clump-like feature at a similar 
 V-band magnitude to the red-end of the horizontal 
 branch (HB) described  below; and we  discuss the clump later in
 Sec.~\ref{redclump}.
   A significant number of  bright red stars occur  
 above the I-band discontinuity that marks the top of 
 the RGB:  They  could be
 intermediate-age AGB stars covering a moderate range of ages and
 metallicities, and are discussed further in Sec.~\ref{youngstars}.
  There is  clear evidence for 
   blue and red horizontal branch (HB) stars at 0.2 $\la$ V-I (mag)
  $\la$ 0.6, characteristic
 of an ancient population of metal poor stars with [Fe/H] $\la$ -1.5\,dex.
 This is the first direct evidence for such a population in this galaxy (also
 see Geisler et al. 1999). 
  There are stars  surrounding the HB at V-I $<$ 0.5\,mag, for example,
 that could comprise main sequence stars, blue loop stars as well as
 a foreground population of stars; and those 
  will be  discussed further using isochrones 
 in Sec.~\ref{youngstars}.\footnote{It is however interesting to compare  the observed  morphology with the prediction of 
 MD99 (their Fig. 16), which is similar}

 In NGC 205, somewhat like NGC 185, there is a  prominent sequence 
 of  RGB/faint-AGB stars,   
  similar to that  presented by Lee (1996). That upper sequence
  does not appear to have a homogeneous colour distribution at a given
 I-band magnitude, which we 
  detail in Sec.~\ref{Secabundance_distrib}.
    We fail to detect evidence for, or against, a  HB 
 in the central region due to the
  magnitude limit of the photometry. However, there is 
 an increased density of stars at  about I = 24.5-25.5, V-I $\la$0.6\,mag, 
  the location  where a HB might be
 expected in outer fields (F2 and F1), because crowding-noise
 is less there, but  
 contamination from the stars in M31's halo is more likely, 
 as detailed later below.  
 There is also a scatter of stars redward
 of the RGB/faint-AGB (V-I $\ga$ 3); they could be 
 very metal-rich stars that belong  to M31 and are briefly
 mentioned later in Sec.~\ref{tails}.
 Lastly, the NGC 205 CMDs  contain several  bright, blue
 stars (at I$\la$23, (V-I)$<$ 1) that are probably
 mostly main sequence and blue loop stars; they are examined further in
 Sec.~\ref{youngstars}.

  For an analysis of how much the NGC 205 CMDs are contaminated by M31 stars, 
  we  used a B-band isopleth map of NGC 205 (with an
 elliptically averaged surface brightness 
  profile of M31 subtracted;  
 Fig. 9 of Ferguson et al. 2002) and   minor axis
 surface brightness data for M31 from Walterbos \& Kennicutt (1988; WK88). 
  It follows that 16-40\% of the brightest stars (probably field
  stars, upper
  MS/blue loop stars and upper AGB/RGB stars) across the  F1 field 
 belong to M31. 
 Adopting that level of contamination, we would implicitly assume 
 the same age and chemical composition for NGC 205 and the halo of M31, 
  and we therefore made a different check: Applying our selection
 criteria, detailed in Sec.~\ref{obs_datared}, 
 to  photometry from  Bellazzini et al. (2003) for 
 a nearby field\footnote{G58 is 
   11.2$^{\prime}$ distant from NGC 205 
  and lies between NGC 205 and M31.} (G58), we found 
 1193 stars   at I $<$ 22.5\,mag.
 Scaling that number for NGC 205, where the M31 halo is 
  approximately 1.5\,mag  arcsec$^{-2}$ fainter, 
 the estimated contamination is in good agreement with the 
  previous estimate of 16-40\%; and  can be applied  for
  AGB/RGB stars  at I $<$ 22.5\,mag.
  Changing now to the F2 field, 
  the contamination level should be below 1\%,  
 which is negligible, and  even less for the remaining fields 
 which are closer to the nucleus of NGC 205. 
 Additional   evidence  that contamination 
  from M31 stars  decreases towards NGC 205 comes from
 the blue plume or upper main sequence population in M31:  
 I.e., compare the 
 blue plume in fields G58, G64 and G108 from BCF03.
 We note that about half a dozen bright, blue stars occur 
 at V-I $<$ 1\,mag and 20$<$ I (mag) $<$ 21 in each NGC 205 field --
  they may well be foreground stars, even though 
  inspection of nearby (control) fields, G58 and G64 
 (see Fig. 5 \& 6 in BCF03), argues for 
 a negligible surface density of foreground  
 stars. Alternately, they might be members of NGC 205, but that
 interpretation is disfavoured in Sec.~\ref{youngstars}. 

  For NGC 185 and NGC 147, the surface density of (unobscured) 
 field stars (and point-like field galaxies)
 along the line of sight is taken to be 65$\pm$8 and 70$\pm$8  stars
 respectively per WFPC2 pointing  at I$\la$ 22.2\,mag and dominated
 by stars redder than about R-I =0.4\,mag (Battinelli \& Demers 2004a,
  2004b). Their surface density is such that they are 
 unlikely to have a significant effect on 
 our distance/abundance  measurements later. As we consider
 the population of MS and blue-loop stars later, we note
  that some contamination
 by field stars at  R-I $<$ 0.4\,mag is possible, but the level should be 
 negligible ($<$$<$ few percent of the field surface densities given above) 
 on average based on an
 inspection of the CMDs from the Battinelli \& Demers wide-field 
 study. The number of 
 contaminating field galaxies was estimated to be 
 23$\pm$5 per  WFPC2 pointing at 20 $<$ I (mag) $<$ 22 from
  the Medium Deep Survey (Griffiths et al. 1994), which is 
 an unlikely source of significant contamination  that
 was not assessed by the  artificial  star tests.

\section{Median abundances   and distances}\label{Sec_dist_metal}
In the absence of star-to-star abundance information from spectroscopic
  measurements,  popular ways to measure  distances and 
 to infer  [Fe/H] values for ancient stars in
 Local Group galaxies have included
 the RGB method, the horizontal branch (HB)
  V-band magnitude,  RR Lyrae stars, and the turn-off magnitude of 
 main sequence stars. We consider the first two of these.
 Such diagnostic tools have, necessarily, been calibrated using ancient 
  stellar populations (galactic globular clusters; GGCs); and are thus
 best applied to a single stellar population of similar  ancient 
 age (e.g. 13\,Gyr).
 Clearly,  our photometry  does not sample such simple stellar populations  
 but for NGC 185
 and NGC 147  there is evidence that their stellar populations are mostly
 ancient, with a small intermediate-age component -- 
 see MD99 (NGC 185) and Han et al. (1997; NGC 147). 
  For NGC 205 however we are less certain 
 because of the possibility of a  significant
 fraction of intermediate-age stars.

 On the reliability of  the absolute magnitude-abundance relation
 for the HB, on which the HB method relies, we recall the relation
  is controversial in the literature (e.g. see Caputo 1997, and references
 therein) due to the dependence of M$_V$(RR Lyrae) on the composition
 and evolutionary phase of the variable star. 
 That leads to  a possible range in [Fe/H] of the order of 0.2\,dex
 for a given absolute V-band magnitude.  Despite this caveat, 
 there is an advantage to using both the HB magnitude method together
 with the RGB method, namely that in 
 stellar systems  dominated by ancient stars with similar 
 abundance levels, one would expect
 an agreement, within uncertainties, between the two methods.
 Based upon the hypothesis that the RGB is dominated by ancient stars
  with similar  abundance levels to those of  the HB stars, we will make
 a comparison of the inferred abundance estimates.

\subsection{Lower [Fe/H] limits}
 In the present paper, when we refer to stellar abundance (or metallicity), 
 we are referring to [Fe/H] as inferred from the  colour
 ((V-I)$_0$) at a certain
 absolute I-band magnitude for ancient stellar populations. In this way,
 we are making an inference of the 
  metallicity of ancient RGB stars, using an  approach
  that can be applied to  
 other stellar systems. A key concern however is the so-called 
 age-metallicity degeneracy in stellar 
 populations comprising ancient and intermediate-age stars. 
 The trouble is that evolved (i.e., post-main sequence) 
 stars that are between about one and  a few times 10$^9$\,yr
  old evolve along
 the blue side of the RGB causing 
 the inferred median (or mean) colour to be bluer than that of the 
  ancient RGB stars. We use the median colour as opposed to the mean value
  as the median is a  more reliable tracer
 of the predominant colour  when 
 the colour distribution is skewed -- otherwise the median and mean  
 values are identical. However, whichever of these two statistics is
 used,  an inferred  (mean/median) [Fe/H] value will under-estimate
  the actual  [Fe/H] of ancient RGB stars, and so our photometric
 abundance is taken as a lower
 limit as is done implicitly in other studies that determine photometric
 abundances (see Grebel, Gallagher \& Harbeck 2003). 
 Photometric abundances are useful because they are available for 
  a wide range of systems, 
  thereby allowing them to be compared. For
 example, see  Grebel, Gallagher \& Harbeck 2003, their section 2.1, for a
 useful  comparison of nearby dwarf galaxy properties.

\subsection{Applied methods and results}\label{Sec_dist_results}
 For the RGB-tip indicator of distance and [Fe/H], one requires 
 a knowledge of the tip I-band magnitude of the RGB, I$_{\rm TRGB}$,
  and its uncertainty.
    In the case of NGC 205, we  estimated   I$_{\rm TRGB}$
 by applying a slightly modified version of the
  edge-detection method described by Sakai et al. (1996). 
  I.e.,  we evaluated the edge-detection response at fine intervals
 ($<$10$^{-3}$\,mag) 
 followed by smoothing using a 0.05\,mag-wide 
 moving-average window and recorded the 
 peak response in a restricted magnitude range (I$<$21\,mag). 
 That was repeated a few hundred times in a 
  bootstrap way\footnote{ 85\%  of the stars are selected randomly and
 the median  value taken. This is repeated a few hundred
 times, followed by a gaussian  fit to the resulting 
 distribution of values. The peak of the fit
 is taken as the average median value while its 
  standard deviation is taken as the uncertainty.}  to  measure the average tip magnitude and its uncertainty.
 For the  inner two fields combined, 
 we found  I$_{\rm TRGB}$ = 20.85$\pm$0.1\,mag where the uncertainty
 includes a conservative error to account for the probable slight 
 bias introduced  by the smoothing process. 
 For NGC 185 and NGC 147,   
 we could not accurately identify the 
 RGB tip magnitude using the edge-detection approach using
  individual fields as there are too few stars, 
 making the RGB tip poorly defined.
  Using the ensemble data for NGC 185, 
  but  restricting the edge detection to I $<$ 20.5, we find I$_{\rm TRGB}$ = 20.4
$\pm$0.1\,mag  which is close to the value from Nowotny et al. (2003; NK03)   --
 one obtains I$_{\rm TRGB}$  = 20.31\,mag for NGC 185 from their data. However, as their
 detection criterion is much better, owing to significantly more stars,
  we adopt their value, which agrees with previous estimates in the literature
  (Lee et al. 1993; MDA98; NGC 185).
 For NGC 147, we deduce that 
 there are too few stars at I$<$ 21\,mag to allow an accurate measurement
 using the edge detection method; and so, we adopt 
  I$_{\rm TRGB}$ = 20.70\,mag taken from the NK03 -- incidentally, that
  I$_{\rm TRGB}$
 value is in good agreement with the estimate by Han et al. (1997) that was
 based on visual inspection of the I-band luminosity, using the  
 same WFPC2 dataset retrieved for the present paper. 
 The  (median) colour
 of the  RGB/faint-AGB  sequence at M$_{\rm I, -3.5}$ could then be 
  measured. Derived (median) (V-I)$_{\rm TRGB, 0}$ data as well as the 
 inferred  
 [Fe/H] and distance modulus values, using the RGB-[Fe/H] relation
 (Lee et al. 1993; see below),
 and the RGB magnitude and colour ranges detailed below are given in
   Table~\ref{distance_comp2}.

 For the V(HB) method of [Fe/H] and distance estimation,  
   the  relevant relations are (a) the 
  M$_{\rm V}$--[Fe/H] relation 
 for the HB at the location of the RR Lyrae star 
 instability strip, i.e., M$_{\rm V}$ = 0.17\,[Fe/H] + 0.82 (Lee et al. 1993)
  and (b) the empirical calibration relating
 the de-reddened RGB V-I colour at M$_{\rm I, 0}$ = -3.5\,mag 
 ((V-I)$_{\rm 0, -3.5}$) to [Fe/H]  i.e.,
 [Fe/H] = -12.64 + 12.61\, (V-I)$_{0, -3.5}$ - 3.33 \,(V-I)$_{0,
    -3.5}^2$ (Lee et al. 1993); the relation is valid for
   -2.2$<$[Fe/H]$<$-0.7\,dex.  We recall that (m-M)$_0$ =  I$_{\rm TRGB}$ 
  + BC$_{\rm I}$ - M$_{\rm bol, TRGB}$ - A$_{\rm I}$, where BC$_{\rm I}$ is the bolometric
 correction to the I-band magnitude, and M$_{\rm bol, TRGB}$ is the bolometric
 magnitude of the RGB tip magnitude.  Following Da Costa \& Armandroff (1990),
   we take  BC$_{\rm I}$ = 0.881 - 0.243\,(V-I)$_{\rm TRGB}$, and  M$_{\rm
 bol, TRGB}$ = -0.19\,[Fe/H] - 3.81. 
     We obtained median  RGB colours and uncertainty values using
  the previously mentioned bootstrap approach
 in the  M$_{\rm I}$  range -3.3 to -3.7\,mag ((V-I)$_{0, -3.5}$)
  and -4.1 to -3.95\,mag ((V-I)$_{\rm TRGB}$).
 Similarly, we measured the HB V-band magnitude 
 and its uncertainty  for stars 
 at 24.5 $<$ V(mag) $<$ 25.7 and 0.2 $<$ V-I (mag) $<$  0.7.
 We then solved for the distance modulus and [Fe/H] in an iterative way 
 that is approximately self-consistent because both relations used 
 are calibrated against ancient simple 
  stellar populations (galactic globular clusters)
 -- the 
 systematic error arising from the [Fe/H] matching tolerance 
 is $\pm$0.025\,dex. 
 In order to estimate the (random) 
 uncertainty in the derived [Fe/H] value, the iterative
 procedure was re-run using V(HB) + $\sigma_{\rm V(HB)}$ and  again using
 V(HB) - $\sigma_{\rm V(HB)}$.  The final [Fe/H] value, the derived distance
  modulus,  the  V(HB) magnitude and their uncertainties are given  in
  Table~\ref{distance_comp} for NGC 185 and NGC 147.  
 Such measurements
 for the F1 field in NGC 205 are precluded by a poorly defined HB.

Comparing our distance (and [Fe/H]) data from  the two 
 methods we applied, there is a
  good agreement  for NGC 185. As we adopted the 
 TRGB I-band magnitude data 
 from  NK03, it is not surprising that 
 the derived distance modulus  is in excellent agreement
 with  their value, 24.04\,mag, also determined using colour information. Our 
 [Fe/H] lower limit for NGC 185,  -1.11$\pm$0.08\,dex,  
 derived using the V(HB) method 
 (Table~\ref{distance_comp}), is  consistent
 with that from  MDA98, as well as
  their report of an outward radial decrease in [Fe/H]; 
 but is insufficient to confirm it. 
 However, there is  a marked  
  discrepancy between the NK03 finding of -0.89\,dex for NGC 185 and
 our value that can be  explained by a mis-match
 between the  foreground extinction adopted by NK03 
 (A$_{\rm V}$=0.48\,mag) and 
 our  value (A$_{\rm V}$=0.57\,mag, from SFD98): That extinction
 difference causes their  value to be systematically more metal-rich
 than our finding, by about 0.2\,dex.

 For NGC 147 the V(HB) method provides a  distance modulus estimate 
 that is  markedly bigger by 0.1\,mag than the value from the RGB method 
 which relies on the TRGB magnitude from  NK03 and is significantly different
 to their
  high signal-to-noise estimate of  24.44\,mag
 based on the RGB tip magnitude method; and could be hinting that 
 the dominant population on the RGB/faint-AGB is not ancient and / or 
 is more metal-rich than the HB stars.
 As expected, our [Fe/H] data   from the RGB method
  is in fine agreement with
 that of Han et al. (1997) and  agrees well with the 
 finding of NK03 (-1.11\,dex) and Mould, Kristian \& Da Costa (1983), each of
 whom also used colour information.

 Our distance modulus calculation for  NGC 205 leads to (m-M)$_0$=24.76$\pm$0.1\,mag 
 (for E(B-V) = 0.11\,mag) which would place NGC 205 farther behind M31 than
 previously estimated in the 
 literature, e.g. Grebel et al. (2003) and references therein.
 Lastly, our inferred  [Fe/H] lower limit is  -1.06$\pm$0.04\,dex from
  the RGB method,  
  which is markedly more metal-poor than the Mould, Kristian \& Da Costa (1984)
 inference that was based on RGB colour information from
  seeing-limited data ([Fe/H] $>$ -0.85$\pm$0.2\,dex).
  Just like the discrepancy we found for  NGC 185 between our [Fe/H] value
  NK03 result, mentioned above, a probable cause 
  is a  systematic under-estimation of foreground extinction  by about
  (A$_{\rm V}$ =) 0.1\,mag by Mould et al.
  (they assumed A$_{\rm V}$ = 0.23\,mag).

Lastly, it is worth remembering that if future reddening maps with 
 higher spatial resolution than that of the SFD98 maps (6.1\,arcmin) 
 find that SFD98 underestimates extinction at 
 our WFPC2 pointings (FoV$\sim$2.5\,arcmin), our inferred 
 photometric abundance data should be shifted to 
 lower abundances values (e.g., 
 using the Lee et al. 1993 [Fe/H]$_{\rm phot}$--RGB
 relation).

\section{Abundance distributions from  colour
 information}\label{Secabundance_distrib} 
 In the previous section, one of the 
 relations we used to infer abundances is the often employed
 empirical RGB-[Fe/H] relation that is calibrated using ancient RGB stars,
 to infer  a 
  photometric abundance  for the stars in each field.
  In the present section, we examine the distribution
 of abundances in each field, and  
  in order to be able to check the effect of a wide age range in a homogeneous
 way,  we use model RGB-[Fe/H] relations by employing
 a code to generate synthetic CMDs\footnote{Using software from G. Bertelli,
 and  
 based on a library of stellar evolution tracks (Bertelli et al. 1994). It has
 been used in numerous studies in the literature, 
 including Butler, Martinez-Delgado \& Brandner (2004).}. 
  As one does not know the actual age distribution of the stars, we adopt
 a single, mid-range age (10\,Gyr), as Worthey et al. (2004) did, and an
 adequately wide abundance range 0.0001 $\le$ Z $\le$ 0.009 ($\sim$ -2.3 to 
-0.30\,dex)\footnote{Taking [Fe/H] = Log(Z/Z$_\odot$) where Z$_\odot$ =
 0.018.} -- spreading age uniformly in a wide range (e.g. 5-15\,Gyr) 
 causes the inferred abundance distribution to widen
  by about $\pm$0.2\,dex, as expected  (see Worthey 1994; his Table 6). 
  Incidentally, that age dependence is why the abundances derived later
  using the synthetic 10\,Gyr models (see 
 Table~\ref{abund_summary})   produce [Fe/H] values that are
  systematically more metal-rich by about 0.1\,dex than
 the mean data derived in Sec.~\ref{Sec_dist_metal}
  using GGC stars that are a few times 10$^9$\,yr older
 (see Tables~\ref{distance_comp2} and ~\ref{distance_comp}).

 The basic idea here is to  assign an abundance value to measured
  giant star colours
 from a synthetic CMD (10\,Gyr, with random Z values from 0.0001 to  0.009)
  in which realistic photometric scatter is added 
 --  giant stars were selected such that -3.9 $<$ M$_{\rm I}$ (mag) 
  $<$ -3.3 and 0.7  $<$(V-I)$_0$ (mag) $<$ 4.
 Only distributions for two fields in each galaxy are 
 plotted for the sake of clarity  (see Fig.~\ref{abundance_distrib}),
 but  for all fields  we give in Table~\ref{abund_summary}   
   the number of stars per distribution as well as (a) the 
  fraction of stars that are  
 old and metal-poor and or young and  metal-rich ([Fe/H]$<$
 -1.4\,dex)\footnote{That implicitly assumes
 that
 metal-richness is higher on average at successive epochs of 
 star formation}  and (b) the fraction that is metal-rich 
 (-1.4 $<$[Fe/H]$<$ -0.2\,dex); the abundance separation   
  at -1.4\,dex is an ad hoc value roughly mid-way in 
 the range -0.5 to -2.5\,dex.

 For a check of the dependence of the inferred metal-rich tail in NGC 205
   on a significant
 error in extinction (20\%, for instance), owing to its proximity to M31, 
 we made a visual inspection of
 the distribution of stars in the  (M$_{\rm K}$,V-K)$_0$ CMD, shown in
   Fig.~\ref{MK_V_K_cmd}, 
 where the K$^\prime$-band photometry is from Davidge (2003). That photometry
 came from   observations 
  under 0.7arcsec seeing  at r $<$ 1.33$^{\prime}$ (e.g., see his  
  Fig.~\ref{MK_V_K_cmd});
  K$^\prime$-band completeness is above 75\% at K$^\prime$ $\la$
 18.2\,mag and $\sigma_{\rm K^\prime}$ $<$ 0.35\,mag)\footnote{That data
 is  useful in our study at   M$_{\rm K}$ $\la$ -6\,mag only,
   due to a substantial increase in  K-band
 incompleteness at fainter magnitudes.}.
 Accordingly, with the help of the wide dereddened colour baseline (V-K)$_0$, 
  it follows from visual inspection that even with the sizeable hypothetical
 extinction error, the metal-rich component of the RGB 
 is probably at least as rich as -0.7\,dex.

 Another issue is whether there are 
 radial changes in the shapes of inferred distributions, for which one
 can make use of the skewness parameter 
 listed   in Table~\ref{abund_summary}. 
 There is no clear  evidence for NGC 185, but NGC 205  
 becomes systematically redder towards its inner regions.  
 That finding is unlikely to be 
 affected by strong differential extinction, which
 affects only a small fraction ($\sim$ 5\%) of the central field; but there
 remains  the
  possibility of a smaller levels of differential extinction (e.g. 0.05\,mag) 
 over a wider area.  Lastly however, as abundance levels
 could be higher at successive epochs of star formation 
 there remains the possibility of 
  a correlated spatial variation in stellar ages  that
 might act together to give an RGB/faint-AGB colour
  roughly independent of position.

  On the issue of metal-poor stars, it is hard to make a firm statement
 regarding the accuracy of a metal-poor tail reaching to [Fe/H]$<$ -1.5\,dex
    without spectroscopic information and or evidence for a HB, simply 
 because stellar  ages are unknown. 
 For 
 NGC 185 however we know that at least some of the stars in the metal-poor
  tail should
 be ancient because of the presence of a HB reaching to  colours as 
 blue as V-I $\sim$ 0.1 - 0.2\,mag. 

 Another issue is the similarity of the distributions.
   In the case of NGC 205, we determined whether
 the shape of the control field and NGC 205 distributions
 differ. We did this by determining the mean skewness and its uncertainty 
 based on  the bootstrap approach described earlier in 
 Sec.~\ref{Sec_dist_results} -- the derived data is given in
 Table~\ref{abund_summary}. 
 Based on an application of 
 the  Student's T-statistic to the bootstrap samples,
  we find that the  skewness  in the control field (G58)
    differs significantly from that in any of the NGC 205 fields.
  As the skewness in NGC 205 is  unlikely to be the result of 
 significant 
 contamination from M31 (see Sec.~\ref{CMD_description}),
 we infer that the RGB/faint-AGB sequence in NGC 205 is significantly redder
 than that in the control field (G58). The RGB/faint-AGB sequence in NGC 205 is also skewed to redder colours than the sequence in NGC 147 or  NGC 185.
 An exact interpretation for the RGB/faint-AGB sequence NGC 205 is 
 out of reach without modeling using data with a much 
 fainter magnitude limit, but the skewness data hints that 
 a significant fraction of the  RGB/faint-AGB 
 stars occur in a relatively narrow age and or 
 abundance interval compared to the control field.

Finally, we note that
 skewness varies markedly with radial position in NGC 205, in the sense that
 the inner  region  stars are generally 
 redder than the outer region stars; and it is unlikely to be
 fully  attributeable to contamination from M31.

\section{Abundances: Comparison with previous studies and a control field}\label{tails}
 
From the poorly defined 
 red-end of the bright AGB, we infer Z$\sim$  0.1\,Z$_\odot$
 at  (V-I)$_0$ $\sim$ 3.5\,mag, with higher abundances likely for 
 redder colours based on  synthetic CMDs.
   For a comparison, the work of  Gallagher, Hunter \& Mould (1984) suggests
 that the   abundances of the most metal-rich stars  
  in NGC 185 may be less than 1/3 solar (i.e., Z $\la$ 1/3\,Z$_\odot$, = 0.006)
  based on the N$_{II}$/S$_{II}$ ratio for one planetary emission nebula
 near the centre of NGC 185. Similarly,  the mean oxygen 
 abundance (12 + log (O/H) = 8.2)  from several planetary nebulae 
 (PNe) (Richer 
\& Mc Call 1995)  is consistent with a system whose most 
 metal-rich stars have about one third of the  solar abundance.
 Consequently, there is rough agreement with our CMD-based finding.
  
 In NGC 205,  the reddest pair of AGB stars reach to 
  (V-I)$_0$$\sim$4.3\,mag corresponding to Z$\sim$ 0.0065, about one 
 third of the solar value. Consequently, abundances in NGC 205 appear to 
 reach to richer values than in NGC 185, a fact that is supported
 by the  richer mean oxygen 
 abundance for NGC 205 (12 + log (O/H) = 8.6; Richer \& Mc Call 1995).

  Further on the subject of  metal-rich
 stars,  there are some very red stars (V-I $\ga$3) 
 below the I-band RGB tip magnitude in NGC 185 and NGC 205. Whether 
 those stars belong to NGC 205 is unclear, but such stars occur
  in M31 fields. For example, BCF03
 found them in WFPC2 photometry of M31 halo fields 
 and noted that such stars would 
 be highly metal-rich, if they are members of M31.
  Examination of such stars with high
  abundance levels is largely precluded by 
 incompleteness and  model uncertainties.

\section{Age and abundance clues: the red bump and HB in NGC 185}\label{redclump}
 Stellar
 evolution theory predicts that the luminosity  of the RGB and AGB bumps
 in the colour magnitude plane depends on stellar  
 abundance   and  age. Such theory also predicts a red HB 
  in intermediate-age --to--ancient stars. 
  In the present section we consider the red
  bump/clump (RC) and the prominent HB in NGC 185.

 For the mean RC magnitude in each band-pass, we fitted 
 gaussian function to the  RC  a few hundred times in a bootstrap way (see
 Sec.~\ref{Sec_dist_results}) at V-I = 1-1.6\,mag, 
 binned at 0.05\,mag intervals. 
 The mean V(RC), I(RC) and associated standard errors
 are  25.28 $\pm$0.02\,mag and 24.20 $\pm$ 0.01\,mag respectively.
 Consequently, we find that there is no V-band offset between the clump and
 the HB that we determined at  0.2 $<$ V-I (mag) $<$  0.7, 
 within uncertainties -- the offset is 0.01 $\pm$ 0.15\,mag.
 
 It is unclear whether the RC  actually comprises RGB / faint-AGB 
  stars, a red HB, or both. However,
  we may examine the implications of the third option
 using the  model  $\Delta$V$^{\rm HB}_{\rm Bump}$  diagram 
 from Alves \&  Sarajedini (1999) (their Fig. 6).\footnote{It is recognised that  more
 empirical  
 data from AGB populations is required to place the diagram on a firm footing,
 but the theoretical plane is nevertheless taken here 
 to provide clues on abundances and ages.}
 Consequently, based on visual inspection of their Fig. 6 
 we infer that [Fe/H] would be about -1.5\,dex at $\Delta$V(Bump- HB) = 0
 for ancient stars, with a  higher abundance level for intermediate-age stars,
 regardless of whether  the V-band magnitude of the red HB is the same for all
 ages or one considers the  theoretical age-dependence adopted by 
  Alves \&  Sarajedini (1999).

\section{Young  and  intermediate-age stars}\label{youngstars}
\begin{figure}
\includegraphics[height=105mm,width=130mm,angle=0]{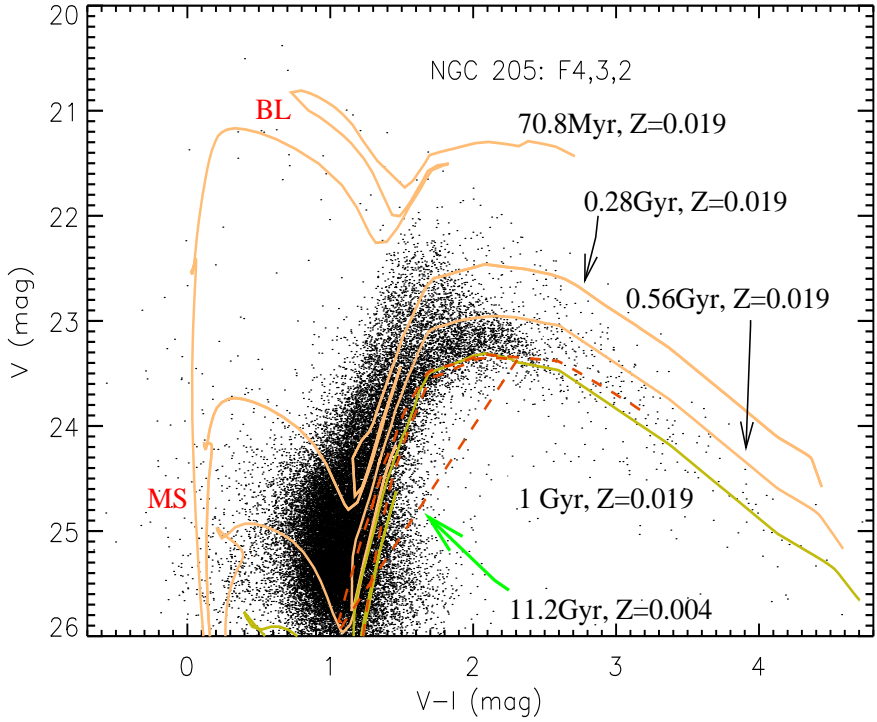}
 \caption{V, V-I CMD for NGC 205 with a set of isochrones overlaid. 
 Stellar mass at the bright-end of the tracks is 
 6.3, 3.6, 2.8, 2.3, and 0.92\,M$_\odot$ respectively, in order of  increasing
 isochrone age. The main sequence (MS) and blue-loop region (BL) of
 the isochrones are marked. 
 See Sec.~\ref{youngstars} for further explanations.}\label{ngc205_V_VIcmd_isochr}
\end{figure}

\begin{figure}
\includegraphics[height=105mm,width=130mm,angle=0]{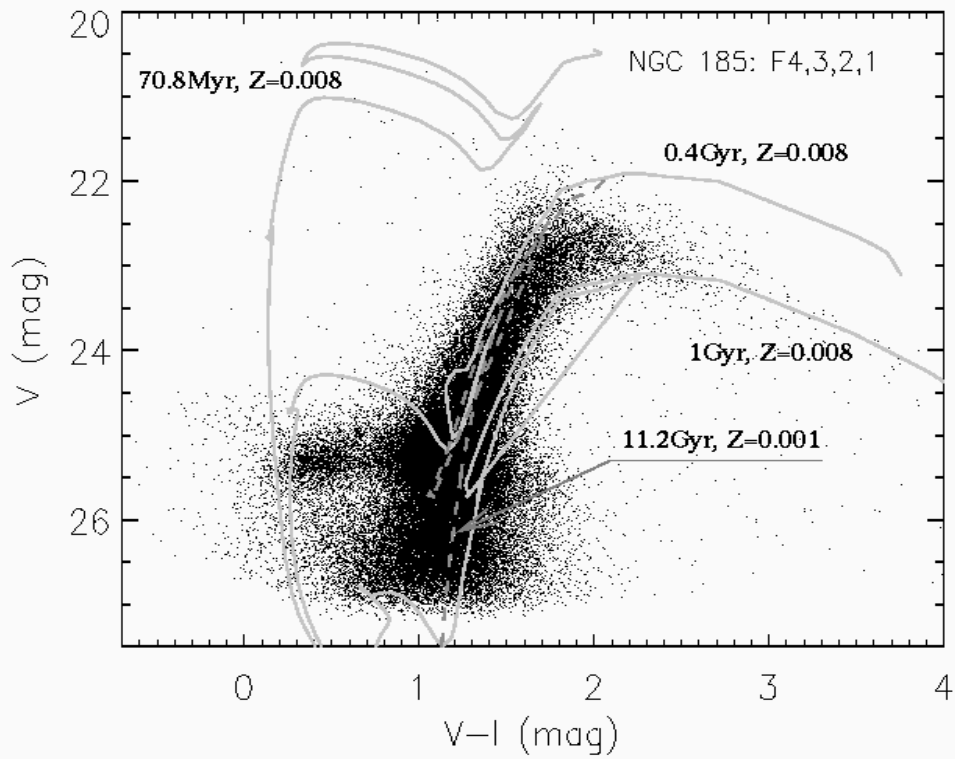}
 \caption{Ensemble V, V-I CMD for NGC 185 with a set of isochrones overlaid. 
 The mass at the bright-end of the isochrone tracks is 
 20.3, 3.0, 2.17, and 0.87\,M$_\odot$ respectively, in order of  increasing
 isochrone age. As explained in the text in Sec.~\ref{youngstars}, the stars
 at V$\sim$ 20.5-23\,mag be field stars. }\label{ngc185_V_VIcmd_isochr}
\end{figure}

\begin{figure}
\includegraphics[height=125mm,width=130mm,angle=0]{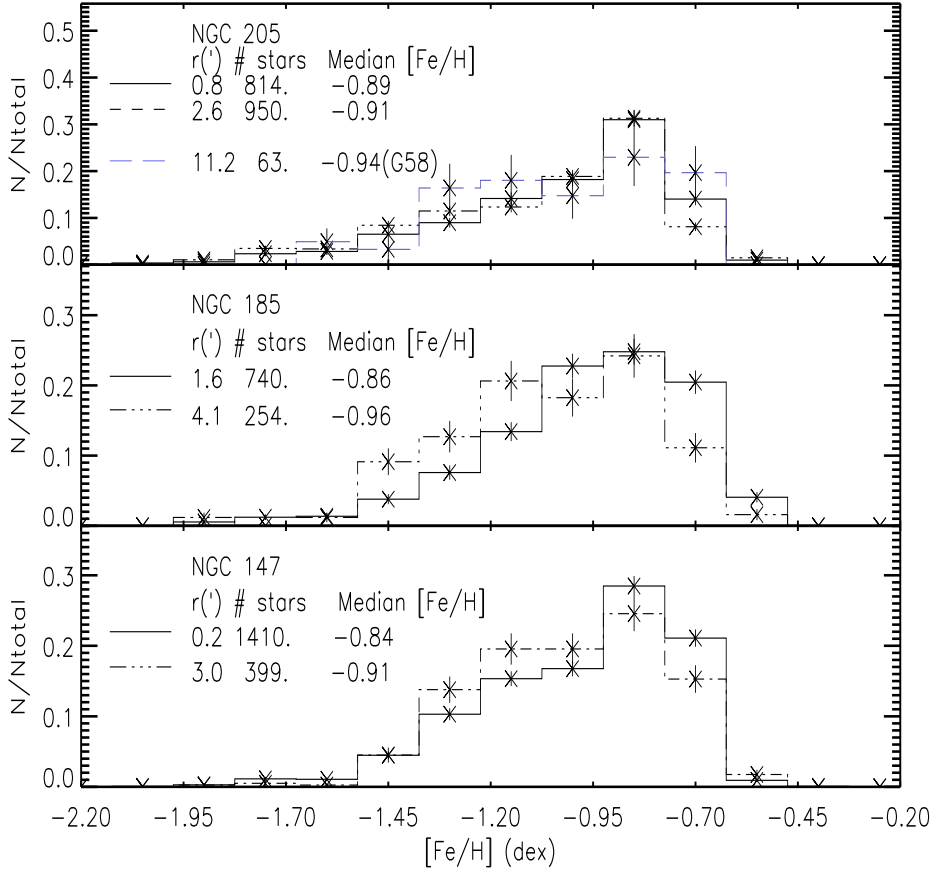}
 \caption{Abundance distributions for NGC 205 (top), NGC 185 (middle), NGC 147
   (bottom) derived by matching colours of bright giants with isochrones 
 of 10\,Gyr old stars.   
 Star count error is $\pm$ $\sqrt{\rm N}$.  We determined the uncertainty in 
 each median [Fe/H] value   using the 
 bootstrap technique outlined in Sec.~\ref{Sec_dist_metal}, which turns out
 to be negligible. See Sec.~\ref{Secabundance_distrib}
  for further explanations.}\label{abundance_distrib}
\end{figure}

\begin{figure}
\includegraphics[width=90mm,height=160mm,angle=0]{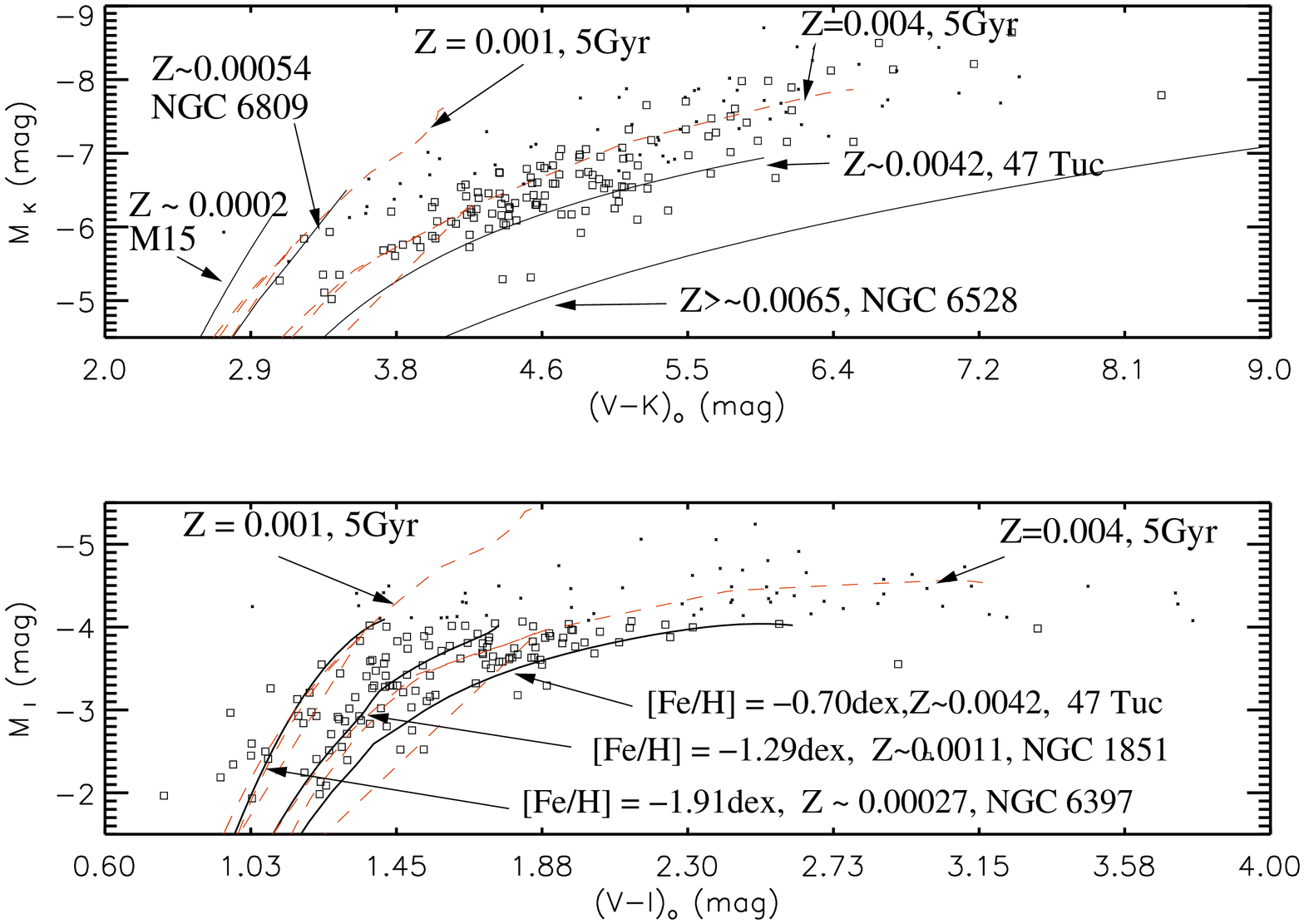} 
 \caption{NGC 205: M$_{\rm K}$, (V-K)$_{\rm 0}$  and
   M$_{\rm I}$, (V-I)$_{\rm 0}$ CMDs for the inner-most pointing
   using (m-M)$_0$ = 24.76\,mag and E(V-I) = 0.18\,mag.
 Stars brighter than the RGB tip magnitude in I-band (dots/tiny squares)
  are separated
 from fainter stars (open squares) for the purpose of presentation. 
  Globular cluster fiducial ridgelines (solid) and stellar evolution
 tracks for the adopted Z = 0.001 and 0.004 Padova models (dashed) are
   overlaid. 
  Globular cluster [Fe/H] data is  on the 
 Caretta \& Gratton  (1997) scale and is listed in 
 Table 1 of Ferraro et al. (1999).
  The [Fe/H] -- Z relation used is 
 [Fe/H]= 1.024\,log\,Z + 1.739, valid for 10 to  15.8\,Gyr-old stars and -2. $<$
 [Fe/H] (dex) $<$ -0.5 (Bertelli et al. 1994;  their Eqn. 11). See
   Sec.~\ref{Secabundance_distrib} for further details.}\label{MK_V_K_cmd}
\end{figure}

  Fig~\ref{ngc205_V_VIcmd_isochr}  and 
 Fig~\ref{ngc185_V_VIcmd_isochr} 
 show the (V, V-I) CMDs for NGC 205 and NGC 185 respectively, in which 
 an arbitrary
 selection of stellar isochrones  from 
 Bertelli et al. (1994) covering young and old ages are
  overlaid  as a visual guide.
 
 The presence of the brightest  stars at V-I $\la$ 0.5\,mag  
 in the NGC 205 CMD is curious
  as one might not suspect 
 a foreground population because of the negligible surface density of such  
 young blue stars in the nearby (control) 
 fields G58 and G64 (see Fig. 5 \& 6 of B03). On the other hand
  they occur in each field with no significant radial trend in the
 dwarf galaxy.  Additionally, 
 from a visual inspection  there is an
  (apparent) paucity of blue stars, e.g.,  V-I$<$0.5\,mag, 
 at  V$\sim$ 21.5-23\,mag that is hard to explain by a conspiracy of
 incompleteness functions, and is inconsistent with a main
 sequence/blue loop population. 
 We take the opinion that the bright, blue stars are not the upper
 part of a main sequence/blue loop stellar population, but rather that
  they are probably foreground stars.

 For the fainter, blue  stars at (V $\la$ 23, V-I $\la$ 0.5\,mag) in
  NGC 205,   which are probably
  dominated by main sequence and blue loop stars, we 
 made a visual comparison  with  several sets of  isochrones, and 
  deduce that
 star formation was probably still active  
 about 1-3 $\times$10$^8$\,yr ago. 
 That finding fits broadly with the discovery that 
   many of the  objects identified as bright stars
  by  Hodge (1973)  are in fact clusters of bright UV sources, 
 two of which
  have stellar ages of 0.5$\times$10$^8$\ and 10$^8$\,yr
  respectively (Cappellari et al. 1999). This confirms
  earlier reports of recent episodes  of star formation
 in the nuclear region, based on IUE\footnote{International 
 Ultraviolet Explorer} observations with a 10$^{\prime\prime}$ 
 $\times$ 20$^{\prime\prime}$ aperture (Wilcots et al. 1990).  
  Furthermore, complementary information comes from 
  near-IR studies of  AGB peak brightness and J-K colours of the
 blue AGB sequence that find  there are  AGB stars
  near the centre of NGC 205 that formed less 
 than 10$^8$\,yr ago (Davidge 2003; his Fig 5).
 Lastly, we note that with an orbital period of 10$^8$\,yr (Cepa \&
  Beckman 1988), it is possible that a past disk crossing 
 triggered the recent star formation activity.

   In the case of NGC 185, there are about 
 two dozen blue stars (V-I $\la$1\,mag)
   at V $\sim$ 20.5-23\,mag in the ensemble CMD presented in 
   Fig~\ref{ngc185_V_VIcmd_isochr}.   
 A comparison with models (see Sec.~\ref{CMD_description}) indicates
 that the bulk of them are probably foreground stars owing to 
  NGC 185's low galactic latitude, but some  
 red supergiants from NGC 185 may be present as noted by
 Nowothny et al. (2003) and Martinez-Delgado \& Aparicio (1998).
  Another point against them as members of NGC 185 is that 
  a significant number of fainter stars would be needed if the
 stars were actually MS and blue loop stars.
  However, the presence of a significant population of 
 faint blue  stars 
 at  V$\ga$24\,mag (excluding the HB of ancient  stars  at
   V$\sim$25-25.5\,mag) argues for a real population of  MS
 and blue loop stars. Guided by the isochrones in
   Fig~\ref{ngc185_V_VIcmd_isochr}, one infers that   
  star formation  was probably still active  about 4$\times$10$^8$\,yr
 ago.

  We have not tried to derive  the 
 star formation history of the faint blue stars at  V$\ga$24\,mag
 due to significant incompleteness at such faint magnitudes.
 However, although  
 we detect the faint population of  blue stars at V$\ga$ 24.5\,mag
 based on visual inspection of the ensemble (V, V-I) CMD, 
   the model produced by MD99 (their Fig. 16), based on a synthetic CMD
 analysis,  appears to be
  markedly fainter.  Thus, the  mean SFR density 
 for the last 10$^9$\,yr is probably above the MD99 estimate, 
 i.e., $>$ 2.6$\times$10$^{-9}$M$_\odot$\,yr$^{-1}$
 pc$^{-2}$ at r $<$ 2$^\prime$.

 Next, we probe whether the mean age of young AGB stars
  varies radially with respect to older stars.
 For that reason
  we list in Table~\ref{AGB_RGBratios} the star counts and count ratios for 
 bright, young AGB stars (I$<$ I$_{\rm TRGB}$) and stars on the upper
  RGB/faint-AGB 
 sequence, defined in colour--magnitude space by boxes A and B 
respectively in Fig.~\ref{cmds1} for NGC 185; we shifted the boxes 
 appropriately for
 NGC 205 and NGC 147 with regard to distance and extinction differences.
  For NGC 205, it follows that a real
  increase occurs in the mean ages of bright, young AGB stars with respect to
  older stars  towards the nuclear region, 
  even if a conservative contamination level of 40\% in the F1 field is 
 included.   Simulating conservative systematic errors 
 of 50\% in E(V-I) and / or  0.1\,mag, for example,
 in the distance modulus causes  a shift of a 
 few percent in bright, young AGB star 
 counts, but the conclusion remains unchanged. 

  In the  NGC 185  
 (and NGC 147) data, we find no statistically significant field-to-field
 changes in   the  mean
 ages of bright, young AGB stars with respect to older stars.
 Due to a relatively
 short crossing time in and between the WFPC2 fields, the  AGB and older
 stars could be  well-mixed   (e.g. conservatively, 
 take an rms velocity of 1\,km\,s$^{-1}$), a case that
  would be compatible with 
 the apparent absence of field-to-field variations.

\subsection{Do the intermediate-age populations
 in NGC 185, 205 and 147 differ ?}
 We can make a stab at answering
 the question using Table~\ref{AGB_RGBratios} which lists
  the  number of AGB stars that are
 relatively young (i.e., are above
 the RGB I-band tip magnitude in (I,V-I) CMDs),  normalised
 by the number of fainter (and mostly older) stars, selected from
 suitable colour magnitude boxes (see Fig.~\ref{cmds1}). 
 In fact, we are using  (i) the fact that the   brightness 
 of a young AGB star of a given chemical composition
 is a function of its age, meaning that brighter AGB stars are
 younger and (ii)   and that the  population of 
 ancient stars, which is probably significant,
  is  fainter than the RGB I-band tip magnitude. 
  We assume that there is a negligible
  contamination from (a) RGB stars near the 
  RGB tip magnitude that might arise due to photometric scatter
  and (b) the  number of 
 Mira variable stars occuring near the tip of the 
 RGB that are in their bright state. 
 As the fractional contributions of faint-AGB and RGB stars is unknown,
 we consider the available low time resolution (wide-age
  bins) information offered by our star count data.
 Consequently,  we infer that 
  the contribution of stars  that formed in the 
  past  several 10$^{9}$\,yr
  is significantly greater in NGC 205 than in NGC 185 or
 NGC 147.  For  NGC 147 the contribution 
  is significantly greater than that in NGC 185.

\section{On the nuclear star cluster in NGC 205} \label{NC_SBfitting}
NGC 205 offers an opportunity to 
 study a bona fide nucleated  dwarf satellite in our neighbourhood. We 
 perform a detailed assessment of a number of  
 the nuclear cluster (NC) properties, namely colour, size and  intensity.

\subsection{Profiling method}
For a determination of an NC's structural properties there are
 a number of useful studies to refer to. For example, 
 see B\"oker et al. 2004 and  references therein (also see Walcher et
 al. 2005). 
  We  fitted a separate analytical profile to the NC and to
 the underlying galaxy surface brightness profile that we model as a 
  S\'ersic profile;  such a  profile
   is regarded as plausible based on galaxy morphology studies and 
 has the form  I$^{\rm gal}$ (r) =I$_{\rm
  0}$$^{\rm gal}$
 exp\,[(r/r$_{0}$)$^{\rm 1/n}$], where a S\'ersic index n = 1
 represents an exponential profile and n = 4 is a de Vaucouleurs law. 
  As the radial coverage of the central WFPC2 field
  makes an accurate fit to the  galaxy's profile 
  hard, we fitted the S\'ersic profile to 
 the R-band surface brightness profile from Peletier (1993) 
 by trial and error
 at r = 20$^{\prime\prime}$ to 250$^{\prime\prime}$. For the V-band data, 
 that Peletier profile was  simply shifted to match the V-band WFPC2/PC1 data at  r
 = 9$^{\prime\prime}$. At such distances from the NC, the light 
 contribution from the NC  is negligible, as is the 
 effect  of the HST WFPC2 PSF on the fitting.

\begin{figure*}
\includegraphics[width=175mm,height=75mm,angle=0]{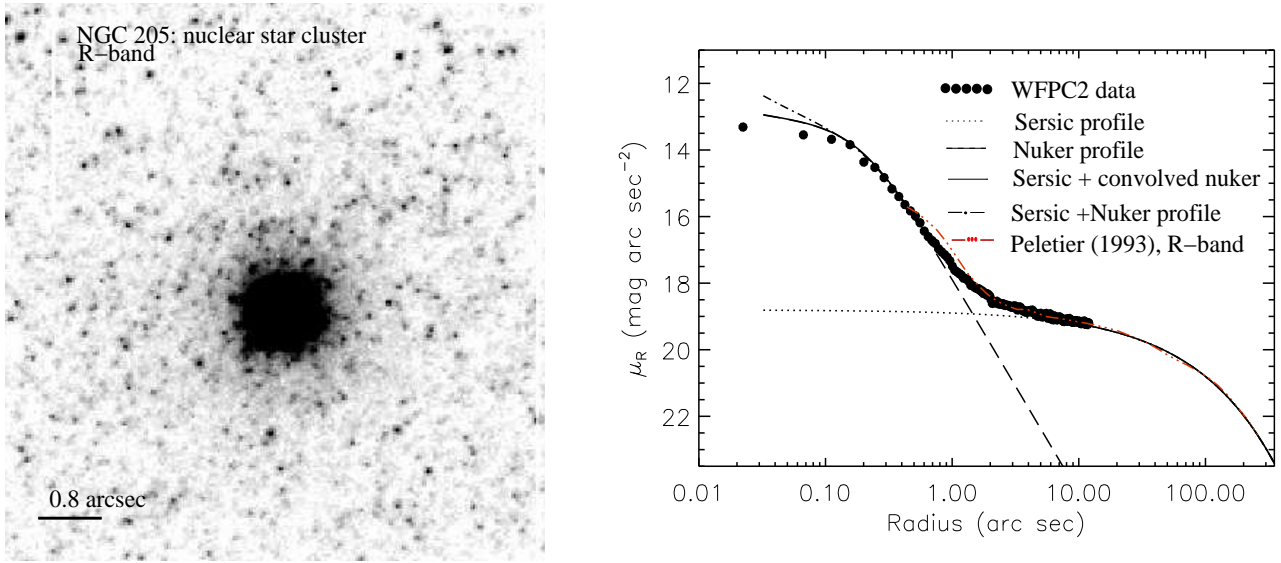}
 \caption{Left: R-band image of the central region of NGC 205.
 Right:  radial surface brightness plot of WFPC2 R-band data and derived 
 surface brightness profiles.  
}\label{ngc205_cntr_cluster}
\end{figure*}

  To model the intrinsic cluster profile,  
 we  used the nuker-law profile  (Lauer et al. 1995)  that is 
 implemented in GALFIT (Peng et al. 2002).    The
 `best' model parameters are those that minimize
  the deviation from the data in a $\chi^2$ sense,   
  and is determined by minimizing residuals
 between the model and the original image. 
 We  note that the nuker model is a fine choice 
 for the present study as it provides the best fit to the cluster data
 in a   reduced-$\chi^2$ sense,  so that  reliable magnitudes and colours
 may be derived.  We used GALFIT in a circular
  aperture (r $<$ 4$^{\prime\prime}$), centred on the NC, 
 that is dominated by emission from the NC; the underlying
  galaxy profile had been subtracted beforehand. Significantly, 
 the GALFIT software takes the input instrumental PSF into account
  by convolving  the analytical cluster profile with it. We made
 a PSF  for each band-pass using
 isolated stars,  modeled  by a moffat function with power index of 2.5. 
 A technical point is that the width of a WFPC2 stellar PSF for a given filter 
  depends on the spectral type(s) of the reference stars; 
 however, as the cluster is well-resolved   
 a mismatch between the spectral energy distribution of the PSF and the 
 NC   has a negligible effect on the derived  intrinsic
 profile of the cluster.  Circular symmetry was not assumed during cluster
  profile fitting which  resulted in a minor-to-major 
 axis ratio of about 0.92, similar to q=0.90 $\pm$0.05 
from Capellari et al. (1999),  q = 0.89
 measured by Heath Jones et al. (1996) using PC1/F555W data, all based 
 on data of comparable angular resolution that resolve the NC well. 
 The actual R-band profile, the fitted and derived intrinsic profiles
 as well as the Peletier (1993) profile are included in
  Fig.~\ref{ngc205_cntr_cluster}.

\subsection{Effective radius and intensity}

In order to quantify the cluster size, we used the so-called
 effective radius, r$_{\rm e}$, the radius of  
 an area that contains half of
the cluster light in projection (see Carlson \& Holtzman 2001); and  
 deduced it for each band-pass  from  the
 derived intrinsic profile of the NC.

To check how much the measured r$_{\rm e}$ (and intensity) 
 depends upon small errors in the 
 wings of the fitted NC profile, we  masked 
 15\% of the PC1 pixels, randomly selected, and fitted the nuker profile. 
 After repeating this several times, we determined  the mean and 
 standard deviation for each of the affected 
 parameters (M$^{\rm C}_{\rm V/R}$; r$_{\rm e, V/R}$; log\,I$_{\rm e, V/R}$) (see Table~\ref{NCproperties}).  

 Taking the derived (dereddened)
  colour and size at face value, we conclude that 
 the NC's blue and red stellar components have similar
 compactness, as judged by the FWHM\footnote{The azimuthally-averaged 
  FWHM are 0.276$\pm$0.007$^{\prime\prime}$ (R) and 0.258$\pm$0.007$^{\prime\prime}$ (V).} parameter, 
 but there is an excess of blue light with respect to red
 in the cluster's outer region -- the measured NC's  (apparent)
  effective radius is smaller by a factor of 
 2.4 in R-band than in V-band.
  Some tentative supporting evidence for
  a blue excess comes  from  Cappellari et al. (1999) 
 who noted that the NC looks
   more elongated in the UV than at longer wavelengths (e.g V-band).
 However,  caution is needed regarding our interpretation
  because the possibility
 of  significant differential extinction,  on a length-scale 
 similar to the cluster size, remains.

 We find that like the NCs in late-type spirals, which  have 
  10$^6$-10$^8$\,L$_{\odot, \rm I}$ (deduced 
 from B\"oker et al. 2002), and unlike GGCs, the NC in NGC 205 
  is quite bright, 
 with 10$^{6}$\,L${_\odot,\rm R}$ (10$^{\rm 0.4(M_{\rm \odot,R} -
  M_{\rm R, 0}^{\rm C})}$) where M$_{\rm \odot,R}$ = 4.31\,mag and
  M$_{\rm R, 0}^{\rm C}$ is given in Table~\ref{NCproperties}.
 
Lastly, it is  worth bearing mind that
 the NC may be a multi-component population, as suggested by the (apparent)
  blue excess in its outer region, possibly a merged pair of
 globular clusters (but also read the Walcher et al. (2005) discourse
 on  possible NC formation scenerios).
  In that case the youngest/bluest component
 would probably be younger than
 the mean age derived from the integrated V-R colour (see the next section), 
 and could therefore be less than a few times 10$^8$\,yr old.

\subsection{Mean age}
  Clues regarding mean age can be drawn from the NC's colour. 
  Mean age can also be estimated from the clusters luminosity and
  mass-to-light ratio; however both of these require a mass estimate, whose 
  significance would be uncertain.

 The NC in NGC 205 is bluer 
 than catalogued  Milky Way globular clusters  (Harris 1996), whose
 bluest clusters reach to V-R $\sim$ 0.4\,mag. Consequently, 
 the NC's colour hints that the NC is not
 ancient, a supposition  that is supported by spectra 
 from Da Costa \& Mould (1988), especially strong hydrogen lines, that
 are similar to those seen in the spectra of young
 globular clusters in the Magellanic Clouds.

 A further way to  get a handle on the mean NC age is through a 
 comparison of the cluster colour with model predictions, again for a single
 burst of star formation. 
  Based on models of an instantaneous
 burst of star formation that produces a 10$^6$M$_\odot$
 stellar cluster\footnote{Taking an IMF slope of 2.35 or 3.3; a lower mass
 cutoff of 1\,M$_\odot$;  an upper mass cutoff of 30 or 100\,M$_\odot$; and Z
 = 0.001 - 0.040\,dex}, a 
 (dereddened) V-R index above 0.3\,mag would only be expected for 
 an approximately 0.5\,Gyr-old cluster 
  progressing to values as blue at (V-R)$_0$ $\sim$-0.1\,mag  at younger ages
 for a wide abundance range, 0.04 $\le$ Z $\le$ 0.001, (Leitherer et al. 1999).
 Consequently,  taking the measured mean colour  at face value and assuming
 a single  epoch of star formation,  one 
 would argue for  a young system up to a few times 10$^{8}$\,yr old.

\subsection{Discussion -- Cluster size}
 Two key issues will be examined briefly here. They  are:
 (a) Is it significant  that 
 the NC in NGC 205 is two-to-three times smaller than those in the 
  Virgo sample of  Geha, Guhathakurta \&  van der Marel (2002) ?, and (b) Are
  typical galactic globular clusters akin to NCs ?
 
 On the first  question, 
 it is interesting that the NC in six bright dE galaxies in the Virgo
 cluster  examined by  Geha et al.
  have  V-band effective radii of 8-13\,pc, larger than the NC in NGC 205
 (r$_{\rm e, V}$ $\sim$ 4\,pc). 
 However, they sampled the bright end of the dE luminosity function.
 Consequently,  future studies extending to the NCs of fainter 
 Virgo dEs should aim to state 
 whether or not the observed sizes are ubiquitous in the  Virgo cluster.

 On the link, if any, between  nuclear star clusters and 
 galactic globular clusters, which are non-central objects, it is
  of at least academic interest to make a brief comparison.
 We note  that the  
 median  r$_{\rm e}$  of GGCs  is almost 3\,pc, with the bulk of them below
 10\,pc (see  B\"oker et al. 2004; Fig. 4). 
 So, although similar to some GGCs in terms of size, GGCs 
 are ancient unlike the NC in NGC 205 which may be up to a 
  few times 10$^8$\,yr old as surmised in the previous sub-section.
 Consequently,  the fact that the average 
 size of the NC in NGC 205 ($\sim$ 3\,pc) happens to  be compatible  with a
 massive  GGC  might be hinting that NC size is not strongly age dependent.

 \section{Has  NGC 205 experienced tidal stripping ? }\label{stripping}
That NGC 205  may have experienced interactions with M31
   has been suggested in the literature  (e.g., velocity dispersion:  Bender,
   Paquest  \& Nieto 1991; twisted outer isophotes: Choi, Guhathakurta \&
   Johnston   2002).  Interactions can cause  tidal stripping of
  stellar matter that could produce stellar sub-structure in 
 so-called  streams. The most prominent 
 sub-structure around M31 belongs to its giant stellar stream  (Ibata et
al. 2001; Ferguson et al. 2002). Recent kinematical evidence plus modeling
 suggests that it  probably does not include NGC 205  (Ibata et al. 2004).
  Our distance modulus 
 measurement for  NGC 205, (m-M)$_{\rm 0}$ = 24.76$\pm$0.1\,mag (when E(B-V) =
   0.11\,mag),  
  would  support that
 result by  placing NGC 205 at d $\sim$ 890\,kpc,   putting
  it  farther behind M31 than previously estimated.

\section{Conclusions}\label{Conclusions}
We presented the first detailed 
 study of NGC 205 and NGC 185 using
  WFPC2 data. Our key results are:

\vspace{0.5cm}

\noindent {\bf NGC 185} 
\begin{itemize}
\item  There is an ancient stellar sub-population in NGC 185 with [Fe/H] $\la$
  -1.5\,dex,  based on the presence of a well-defined horizontal branch.

\item We inferred the median [Fe/H] of ancient stars ($>$10\,Gyr) in 
 NGC 185 from
 colour information to be $>$ -1.11$\pm$0.08\,dex, assuming E(B-V) = 0.184\,mag.

\item In NGC 185, 
 we found  no V-band offset between the red bump/clump and
 the HB V-band magnitude.  From a comparison with theory, that finding
 suggests that
 ancient stars have [Fe/H] $\sim$ -1.5\,dex,  
 with a  higher abundance level for intermediate-age stars (1-10\,Gyr).

\item   We find that star formation was probably 
 still active about 4$\times$10$^8$\,yr ago. 
\end{itemize}

\noindent {\bf NGC 205}  
\begin{itemize}
\item Whether  there is a  horizontal branch 
  in NGC 205 is still an open question as the 
 photometry's limiting magnitude is insufficient to 
  detect central HB stars if they are present. 

\item We determined that the RGB/faint-AGB at each pointing in 
 NGC 205 is significantly
 skewed to redder colours than that of our control field.
  That
is probably the result of a narrower range in stellar 
 abundances and or age for a significant fraction of the stars in NGC 205
 than in the control field located on the outskirts of M31.

\item We inferred the median [Fe/H] of ancient stars in 
 NGC 205 from
 colour information to be $>$ -1.06$\pm$0.04 \,dex,
  assuming E(B-V) =  0.11\,mag. 

\item  Our analysis indicates 
 that star formation  activity in NGC 205 
 was  probably still active  less than 3$\times$10$^8$\,yr ago.
 Such recent star formation activity 
 could have been triggered by an interaction with M31.

\item  We derived a new distance estimate using the RGB I-band 
  tip magnitude method, obtaining (m-M)$_0$ = 24.76$\pm$0.1\,mag, adopting
  E(B-V) = 0.11\,mag.

\item  Several  properties of the nuclear star
  cluster in  NGC 205 have been examined:  In terms of size
 and intensity, it is 
 like a galactic globular cluster or  a nuclear cluster 
 in a late-type spiral galaxy;
  but from a comparison with models, the blue colour  hints
 that its stellar population is young, up a few times 10$^8$\,yr old.
  The (apparent) V-/R-band effective radii indicates  
 an excess of blue light with respect to red light
 in the cluster's outer region. 

  Like the NCs in late-type spirals, which  have 
  10$^6$-10$^8$\,L$_{\odot, \rm I}$ (deduced 
 from B\"oker et al. 2002), and unlike GGCs, the NC in NGC 205 
  is quite bright, with 10$^{6}$\,L$_{\odot,\rm R}$.
\end{itemize}

\subsection{Conclusion -- a brief comparison of NGCs 205, 185 \& 147}   \label{compare_gals}   
In the introductory section  
 we raised key questions  
 about the similarities and differences between
 the three M31 dwarf companions.
Although an investigation of whether the galaxies had significant star
formation episodes since the primeval occurrence is precluded  
  by the low temporal resolution afforded by our CMDs and also  
 by the bright limiting magnitude of our CMDs, we can compare  
 the galaxies in other ways, namely via the abundances of certain
 sub-populations; the most recent epoch of star formation; and 
 the contribution of intermediate-age stars to the total population.

  NGCs 205, 185 and 147 do not show  marked differences 
 in  their inferred median photometric abundances (from colour information)
 for ancient stars, which 
  correspond to  intermediate or richer values
 ([Fe/H] $>$ -1.1\,dex).
  However, both NGC 147 and NGC 185 have a sub-population of ancient stars with
 [Fe/H] $\la$ -1.5\,dex, based on the presence of a HB in each of them.
 The difference between these [Fe/H] estimates (i.e., 
 from the RGB colour and HB
 methods)   could be explained if each contains an  ancient and or  
 intermediate-age population  of predominantly metal-rich stars with
 [Fe/H] $\ga$-1.5\,dex. 
  Alternatively, the discrepancy
  may be resolved to some extent if extinction has been under-estimated.
 For example, based on the RGB colour--[Fe/H] relation  from Lee et
 al. (1993), an under-estimation of our NGC 185 E(B-V) data
  by  0.04\,mag, for example, 
 leads to a  more metal-rich abundance ([Fe/H]) by about 0.2\,dex.
 Accordingly, we conclude that the ancient stellar populations of the 
 three galaxies appear to have broadly similar [Fe/H] values; but these need 
 to be scrutinized by detailed future star formation history analyses
 and spectroscopy-based studies.
 
 On the issue of the most recent epoch of star formation activity, 
  NGC 185 and NGC 147 differ markedly, in 
 the sense that star formation occured more
  recently in NGC 185 ($\sim$  4$\times$10$^8$\,yr ago) than in NGC
  147 ($\ga$ 1\,Gyr ago; Han et al. 1997).
  This is a restatement of the known crisis 
 in explaining the differences in the NGC 147/NGC 185 pair (Mateo 1998).  
 For example, if 
  gas origin in dwarf galaxies is largely internal (via dying stars),
  and they are still capable of   
 gas accumulation at the present epoch, there is the  suggestion
 (Mateo 1998; his Sec. 4) from the lack of gas and recent
 star formation activity that NGC 147 has avoided accumulating 
 gas since the last star formation episode(s).
 
 For the intermediate-age population  we probed
 whether their contribution  varied significantly among the three galaxies.
 We inferred from an AGB--RGB star count analysis that 
  the contribution from stars  that formed in the 
  past  several 10$^{9}$\,yr
  is significantly greater in NGC 205 than in NGC 185 or
 NGC 147.  For  NGC 147 the contribution 
  is significantly greater than that in NGC 185.
 As NGC 147 and NGC 185 are similar in terms of mass and size,  
  an implication  is that  NGC 147
 has had a markedly 
 higher star formation rate at intermediate-age epochs, compared 
 to that at older epochs, than in NGC 185.

\acknowledgments
 T. Davidge and M. Bellazzini are  thanked for the 
  photometry data they provided, as is F. Ferraro for providing
 globular cluster RGB ridge-line data.
  It is a pleasure to thank  J. Walcher for useful comments
 on an early draft.  D. Gouliermis is thanked for a pertinent 
  comment on young star clusters and the anonymous referee 
 is warmly 
 thanked for a careful reading of the manuscript and valuable comments.
DJB devotes this work to the memory of his brother Martin Mary de
 Burgo Butler.

\clearpage

\clearpage

\begin{deluxetable}{lllccc}
\tabletypesize{\scriptsize}
\tablecaption{Source information}
 \tablewidth{0pt}
\tablehead{
  \colhead{Object/} &   \colhead{$\Delta$ r} & \colhead{Filters}   & \colhead{No. of}  & \colhead{Integration}  &   \colhead{No. of stars} \cr
\colhead{Field} &  \colhead{}     &  \colhead{}      &   \colhead{Frames} & \colhead{Time}   & \colhead{}     \cr
  &  \colhead{($^{\prime}$)}     & \colhead{}    &    &     \colhead{(s)} &    
} 
\startdata
 (0)  & (1) & (2) & (3) & (4) & (5) \\
  \hline
\rm NGC 185 & &  & & & \\ 
 F4  & 1.6   &  V/I & 2/3 & 1300/2800  &   22857 \\
   F3 & 2.1      &  V/I & 2/3 & 1300/2800 &  22833 \\
   F2 & 3.3      &  V/I & 2/3 & 1300/2800 &  20747 \\
   F1 & 4.1      &  V/I & 2/3 & 1300/2800 & 16899 \\
 \hline
\rm  NGC 205  & & &  &  \\
  F4 & 0.3      &  V/I & 2/3 & 1300/2700  & 8957 \\
  F3 & 0.8     &  V/I & 2/3 & 1300/2700 &   15718 \\
  F2 & 2.6    &  V/I & 2/3 & 1300/2700 &  24073  \\
  F1 & 8.1    &  V/I & 2/3 & 1300/2700 & 19053 \\
  \hline
\rm  NGC 205   & 0.5    &  V/R & 4/3 & 400/50 &  - \\    
 \enddata
\\
\noindent {(0) Galaxy name; (1) Angular separation of the field taken from MAST, the multi-mission archive 
at STScI, from the galaxy's center to the coordinates listed in the data 
pointings table (nominally the middle of the field of view);
 (3) \& (4) selected filters and frame 
 integration time; (5) Number of stars detected;  `-' Means that stellar photometry was not performed. }\label{obs_log}
\end{deluxetable}

\clearpage

\clearpage

\clearpage

\clearpage
\begin{deluxetable}{ccccc}
\tabletypesize{\scriptsize}
\tablecaption{Adopted extinction  values and galactic coordinates}
 \tablewidth{0pt}
\tablehead{
  \colhead{Object} &   \colhead{A$_{\rm V}$} & \colhead{A$_{\rm I}$}  & \colhead{l}  & \colhead{b}  \cr
 &  \colhead{(mag)} &   \colhead{(mag)} &  \colhead{(deg)$^{\rm a}$} &  \colhead{(deg)$^{\rm a}$} }  
\startdata
 (0)  & (1) & (2) & (3) & (4)  \cr
  \hline
NGC 185    & 0.570  &   0.273  & 120.79 & -14.48  \cr 
NGC 205    & 0.34$^{\rm b}$ &  0.16$^{\rm b}$  & 120.72 & -21.14  \cr
NGC 147    & 0.535  &   0.257  & 119.82 & -14.25  \cr 
 \enddata
\\
\noindent {\rm Extinction values from Schlegel, Finkbeiner \& Davis (1998);  $^{\rm  a}$: Galactic coordinates are from the Simbad database; $^{\rm b}$:  (see
Sec.~\ref{Sec_extinction} for further explanations). }\label{Extinction1}
\end{deluxetable}

\clearpage

\begin{deluxetable}{lcccc}
\tabletypesize{\scriptsize}
\tablecaption{Derived and inferred data from the  RGB method for each galaxy}
 \tablewidth{0pt}
\tablehead{
\colhead{NGC} &   \colhead{Angular} & \colhead{(V-I)$_{\rm TRGB, 0}$}  &
\colhead{(m-M)$_0$}   & \colhead{[Fe/H]}  \cr
 & \colhead{Separation} & \colhead{$\pm$ r} &  \colhead{$\pm$ r}  &  \colhead{$\pm$ r}  \cr  
\colhead{} &   \colhead{(arc min)} & \colhead{(mag)}  & \colhead{(mag)}  &
\colhead{(dex)} }  
\startdata
 (0) & (1) &  (2) & (3) & (4)  \cr
  \hline
185 &  4.1  &  1.72$\pm$0.01  & 24.10 $\pm$ 0.04  &   -1.12 $\pm$ 0.08 \cr 
        &  3.3  & 1.79$\pm$0.02  & 24.06  $\pm$ 0.04   & -1.20 $\pm$ 0.06 \cr
        &  2.1  & 1.74$\pm$0.04 &   24.09 $\pm$ 0.04  &  -1.11$\pm$ 0.05 \cr
        &  1.6  & 1.99$\pm$0.01 & 24.06  $\pm$ 0.03  &  -1.00$\pm$ 0.04  \cr
        &  Mean$^{*}$  & 1.81$\pm$0.12 & 24.08 $\pm$ 0.02 & -1.11$\pm$ 0.08 \cr
\hline
205 &  8.1  & 1.82$\pm$0.03 & 24.77  $\pm$ 0.03  &  -1.04$\pm$ 0.03  \cr
        &  2.6  & 1.86$\pm$0.03  & 24.76  $\pm$ 0.03   & -1.05 $\pm$ 0.03 \cr
        &  0.8  & 1.89$\pm$0.02 &   24.76 $\pm$ 0.03  &  -1.03$\pm$ 0.03 \cr
        &   0.3  &  1.90$\pm$0.10  & 24.74 $\pm$ 0.05  &   -1.11 $\pm$ 0.03\cr
        &  Mean$^{*}$  & 1.87$\pm$0.04 & 24.76 $\pm$ 0.01 & -1.06$\pm$ 0.04\cr
\hline
147 &  3.0  & 1.85$\pm$0.03 &  24.47 $\pm$0.03 & -1.12 $\pm$ 0.04  \cr
    &  0.2  & 1.82$\pm$0.06 &  24.48$\pm$0.05  &  -1.14 $\pm$ 0.08 \cr
    &  Mean$^{*}$   & 1.84$\pm$0.2 &  24.48$\pm$0.01  &   -1.13 $\pm$ 0.01  \cr
 \enddata
\\
\noindent { \rm Columns: As for Table~\ref{distance_comp}, except for 
(2) Dereddened median colour of the RGB/faint AGB sequence at the absolute
I-band tip magnitude of the RGB (3) Derived dereddened
distance modulus;
 $^{*}$ Mean (and standard deviation) of all values; 
   Error values (r) are  uncertainties in V-I$_{\rm TRGB, 0}$, or  
  the effect of the error in  V-I$_{\rm TRGB, -3.5}$ ([Fe/H]), or 
 the effect of the 
 error in  V-I$_{\rm TRGB, 0}$ and I$_{\rm TRGB}$ ((m-M)$_0$). 
}\label{distance_comp2}
\end{deluxetable}

\clearpage

\begin{deluxetable}{lcccc}
\tabletypesize{\scriptsize}
\tablecaption{Derived and inferred data from the V(HB) method for each galaxy}
 \tablewidth{0pt}
\tablehead{
\colhead{NGC} &   \colhead{Angular} & \colhead{V(HB) $\pm$ r }  &
\colhead{(m-M)$_0$$\pm$ r }   & \colhead{[Fe/H]$\pm$ r}  \cr
 & \colhead{Separation} &  &  \colhead{$\pm$ 0.005} & \colhead{$\pm$ 0.025}
 \cr  
\colhead{} &   \colhead{(arc min)} & \colhead{(mag)}  & \colhead{(mag)}  &  \colhead{(dex)} } 
\startdata
  (0) & (1) &  (2) & (3) & (4)  \cr
  \hline
185 &  4.1  & 25.33$\pm$0.01  & 24.18 $\pm$ 0.00  &   -1.39 $\pm$ 0.03 \cr 
        &  3.3  & 25.24$\pm$0.02  & 24.04  $\pm$ 0.02   & -1.14 $\pm$ 0.06 \cr
        &  2.1  & 25.24$\pm$0.01 &   24.02 $\pm$ 0.00  &  -1.02$\pm$ 0.02 \cr
        &  1.6  & 25.33$\pm$0.01 & 24.11  $\pm$ 0.02  &  -1.04$\pm$ 0.05  \cr
        &  Mean$^{*}$  & 25.29$\pm$0.05 & 24.09 $\pm$ 0.07 & -1.15$\pm$ 0.17 \cr
\hline
147 &  3.0  & 25.57$\pm$0.02 &  24.40$\pm$0.02 & -1.10 $\pm$ 0.03  \cr
    &  0.2  & - &  24.38$^a$$\pm$0.01  &  - 0.98$^a$ $\pm$ 0.02 \cr
    &  Mean$^{*}$   & - &  24.38$\pm$ 0.01  &   -1.04 $\pm$ 0.08  \cr
 \enddata
\\
\noindent  {\rm Columns: (0) Galaxy ID; (1) Angular separation of the field from the galaxy's centre; (2)
 Measured median horizontal branch V-band magnitude; (3) derived dereddened
distance modulus; (4) Inferred lower limit for the
 logarithm of the iron-to-hydrogen 
 abundance relative to the solar value, for ancient RGB stars. 
 $^{*}$ Mean (and standard deviation) of all values; 
   Error values (r) are  uncertainties in V(HB), or indicate the effect of
  that error on (m-M)$_0$ or [Fe/H]. 
 The systematic error estimate for [Fe/H] is 0.025\,dex; it 
 is half of  the  tolerance used  to end the iterative search for  [Fe/H]
 and (m-M)$_0$. Formally, the corresponding
 systematic error in (m-M)$_0$, determined by applying M$_{\rm V}$
 =0.17[Fe/H] +  0.82 from Lee et al. (1993) is 0.005\,mag.
- : Means the HB is poorly defined; 
 $^a$: Adopted the V(HB) value from the outer WFPC2 field in NGC 147.}\label{distance_comp}
\end{deluxetable}

\clearpage

\clearpage

\clearpage

\clearpage

\clearpage

\begin{deluxetable}{lcccc}
\tabletypesize{\scriptsize}
\tablecaption{Star counts and count ratios for
 bright AGB and RGB/faint-AGB stars. See
  Sec.~\ref{Secabundance_distrib} for further details.}
 \tablewidth{0pt}
\tablehead{
\colhead{NGC/} &   \colhead{} & \colhead{Star counts }  &
     \colhead{and}   &  \colhead{count ratios}  \cr
\colhead{region}   & \colhead{F1} & \colhead{F2} &  \colhead{F3}  & \colhead{F4}} 
\startdata
  185  &   & &  &  \\ 
\rm A  &   \rm  15$\pm$4 &  \rm   24$\pm$5 & 47$\pm$7 &   47$\pm$7 \cr
\rm B  &   \rm  1105$\pm$33  &  \rm   1668$\pm$41 & 2193$\pm$47 &  2409$\pm$49 \cr
 \rm A/B  (185) &  \rm    0.013(4) & \rm  0.015(3) & 0.021(4) & 0.020(3)  \cr                            
\hline
 205  &   & &  &  \cr 
\rm A  &   \rm   72$\pm$8 &  \rm    224$\pm$15 &  272$\pm$16  &
219$\pm$15 \cr
\rm B  &   \rm    1706$\pm$41 &  \rm   3554$\pm$60 & 3122$\pm$56 &  2034$\pm$45 \cr
 \rm A/B  (205) &  \rm   0.042(6) & \rm  0.063(5) & 0.087(7) & 0.11(1)  \cr 
\hline
 147  &   & &  &  \\ 
\rm A  &   \rm  69$\pm$8 &  \rm  175$\pm$13 &  & \cr
 \rm B &   \rm    1890$\pm$43  &  \rm  7122$\pm$84 & &   \cr
 \rm A/B  (147) &  \rm   0.037(5) & \rm  0.025(2) &   &  \cr 
 \enddata
\\
\noindent {\rm A: {\rm bright\,\,young\,\, AGB},   B: {\rm upper RGB/faint-AGB}. 
  The  star counts (N) and ratios (corrected for a mean
   incompleteness of about 5\% at 
   I $\la$ 23\,mag).    Star count error
    is $\pm$ $\sqrt{\rm N}$.  The 
    error in the last digit of each A/B value is given in parentheses. 
    Box boundaries are defined in  Fig.~\ref{cmds1}. }\label{AGB_RGBratios}
\end{deluxetable}

\clearpage

\begin{deluxetable}{lcccc}
\tabletypesize{\scriptsize}
\tablecaption{Key abundance distribution quantities}
 \tablewidth{0pt}
\tablehead{
\colhead{Object/} &  \colhead{} & \colhead{} &  \colhead{}  & \colhead{}\cr
\colhead{parameter}   & \colhead{F1} & \colhead{F2} &  \colhead{F3}  
 & \colhead{F4} }
\startdata
 NGC 185  &   & &  &  \cr 
 $\Delta$r ($^{\prime}$) & 4.1 & 3.3 & 2.1& 1.6 \cr
 No. of stars & 254 & 398 & 691 & 740 \cr 
 [Fe/H] (dex) & -0.96  & -0.98 & -0.93 & -0.86 \cr 
  Skewness &  -0.70$\pm$0.01  & -0.57$\pm$0.12 &-0.76$\pm$0.058  &
  -0.98$\pm$0.11  \cr 
 F$_{\rm MP}$ & 0.12$\pm$0.03 &  0.15$\pm$0.03 & 0.11$\pm$0.02 & 0.07$\pm$0.01\cr 
 F$_{\rm MR}$ & 0.88$\pm$0.11 & 0.85$\pm$0.09 & 0.89$\pm$0.07 & 0.93$\pm$0.07 \cr
 \hline
 NGC 205  & &  &  & \\ 
 $\Delta$r ($^{\prime}$) & 8.1 &2.6 & 0.8 & 0.3 \cr 
 No. of stars  & 386  & 950 & 814  &  585  \cr  
[Fe/H] (dex)  & -0.96  & -0.91 & -0.89 & -0.88  \cr  
Skewness &  -0.74$\pm$0.01   & -0.85$\pm$ 0.01 &-1.16$\pm$0.12 &  -1.13$\pm$0.09  \cr 
 F$_{\rm MP}$ & 0.16$\pm$0.03 & 0.16$\pm$0.02  & 0.13$\pm$ 0.02  & 0.14$\pm$ 0.02  \cr  
 F$_{\rm MR}$ &  0.84$\pm$ 0.08 &  0.84$\pm$ 0.06  & 0.87$\pm$ 0.06  & 0.86$\pm$ 0.07 \cr  
\hline
 NGC205/G58$^a$   &   & &  &   \cr  
 $\Delta$r ($^{\prime}$) & 11.2$^b$ & -  & -&  -  \cr 
  No. of stars  & 63 & - & -  & -  \cr 
 [Fe/H] (dex) & -0.94    &  & &   \cr 
Skewness &  -0.62$\pm$0.19   &   & &    \cr 
 F$_{\rm MP}$ &   0.08$\pm$0.047 &  &  &  \cr 
 F$_{\rm MR}$ & 0.92$\pm$  0.24 & &  &  \cr 
 \hline
 NGC 147  &   & &  &  \\ 
 $\Delta$r ($^{\prime}$) & 3.0 & 0.2  & -&  -  \cr
  No. of stars  & 399 & 1410 & -  & -  \cr
 [Fe/H] (dex) & -0.91  & -0.84 & &   \cr 
 Skewness &  -0.56$\pm$0.15 & -0.96$\pm$0.07  & &   \cr 
 F$_{\rm MP}$ & 0.06$\pm$0.01 & 0.07$\pm$0.01 &  &  \cr 
 F$_{\rm MR}$ & 0.94$\pm$  0.1 & 0.93$\pm$ 0.05 &  &  \cr 
 \enddata
\\
\noindent  {\rm `-' means no WFPC2 data available.  The [Fe/H] data refer to median  values. F$_{\rm MP}$ and F$_{\rm MR}$ 
  refer to the fraction of stars that are metal-poor or young 
 ([Fe/H]$<$ -1.4\,dex ) and  the fraction that are metal-rich
 (-1.4 $<$[Fe/H]$<$ -0.2\,dex). $^a$ We applied our photometry selection 
 criteria to photometry from BCF03. $^b$ Angular separation
 from the centre of NGC 205. Star count error is $\pm$$\sqrt{\rm N}$.  For each histogram,  skewness is defined here as 
  1/N\,($\sum_{\rm j=0}^{\rm
 N-1}$((x$_{\rm j}$ - $\bar{\rm x}$)/$\sigma$)$^3$, where x is the abundance
 data, and $\sigma$ is its standard devation. }\label{abund_summary}
\end{deluxetable}

\clearpage



\clearpage

\begin{deluxetable}{lcccccccccc}
\tabletypesize{\scriptsize}
\tablecaption{Summary of derived nuclear cluster properties}
 \tablewidth{0pt}
\tablehead{
\colhead{NGC} & \colhead{d} & \colhead{Disk $\mu_{\rm V}^{\rm 0}$} & \colhead{Disk $\mu_{\rm R}^{\rm 0}$} &
\colhead{m$^{\rm C}_{\rm V}$}& \colhead{m$^{\rm C}_{\rm R}$}& \colhead{M$^{\rm C}_{\rm V, 0}$} & \colhead{M$^{\rm C}_{\rm R, 0}$} &
 \colhead{r$_{\rm e, V}$/r$_{\rm e, R}$} & \colhead{r$_{\rm e, V}$/r$_{\rm e,
     R}$} & \colhead{log\,I$_{\rm    e,\, V}$/log\,I$_{\rm e,\, R}$} \\
          &  &  &  &  &  &  &    &  &  & \\
   & \colhead{(Mpc)} & \colhead{(mag} &  \colhead{(mag} &  \colhead{(mag)} &
            \colhead{(mag)} &  \colhead{(mag)} &  \colhead{(mag)} &   \colhead{(arcsec)} &  \colhead{(pc)} &  \colhead{(L$_\odot$\,pc$^{-2}$)} 
           \\
          &  &  \colhead{arcsec$^{-2}$)}&  \colhead{arcsec$^{-2}$)} &  &
            &  &  &  &  & }
\startdata
  (1)  & (2) & (3) & (4) & (5) & (6) & (7) & (8) &  (9)&  (10)&  (11)  \cr
  \hline 
  205   & 0.89 & 19.22 &   18.82 & 14.49  & 14.32  &
   -10.59  & -10.69  &  0.95/0.43 & 4.3/1.9 & 4.23/4.64    \cr
       & (0.02)  &  &  & (0.04) &  (0.01) & (0.13) &  (0.07)  & (0.02)/(0.03) & (0.2)/(0.25) & (0.01)/(0.05) \cr
 \enddata
\\
\noindent {\rm Notes: Col. (2): Adopted galaxy distance (see Sec.~\ref{Sec_dist_metal}). 
 Col. (3) and (4): Observed  (not inclination-corrected) peak surface 
brightness of the galaxy disk underlying the NC. Col. (5) and (6):  Apparent 
 V- and I-band magnitude of NC. Cols. (7) and (8): Absolute magnitudes of the
 NC corrected for foreground reddening. Uncertainties are given in parentheses
 and arise from fitting errors, a distance
 uncertainty of 0.05\,mag and an adopted 10\% error in foreground extinction.
  Col. (9) Angular effective R-band radius and
 uncertainty as discussed in Sec.~\ref{NC_SBfitting}. Col. (10): Effective
 R-band radius in parsecs. Col. (11): Logarithm of effective R-band intensity
 of the NC, log\,I$_{\rm e,\, R}$ = 0.4(4.31 - M$_{\rm R, 0}^{\rm C}$) -
 log(2\,$\pi$\,r$_{\rm e, R}^2$), and log\,I$_{\rm e,\, V}$ = 0.4(4.83
 -  M$_{\rm V, 0}^{\rm C}$) - log(2\,$\pi$\,r$_{\rm e, V}^2$)  with 
 r$_{\rm e}$ in parsecs.  }\label{NCproperties}
\end{deluxetable}





\end{document}